\pdfoutput=1
\documentclass[11pt]{article}
\usepackage{amssymb}
\usepackage{graphicx}
\parskip \baselineskip
\newcommand{\T}{{\cal T}}
\newcommand{\s}{\subseteq}

\newcommand{\R}{\mathbb{R}}

\newcommand{\ol}{\overline}

\newcommand{\ra}{\rightarrow}
\newcommand{\SC}{{\cal S}{\cal C}}
\newcommand{\Ra}{\Rightarrow}

\begin{document}

\title{Compression with wildcards:
   Abstract simplicial complexes}

\author{Marcel Wild}

\maketitle

{\sl Abstract.}  Despite the more handy terminology of abstract simplicial complexes $\SC$, in its core this article  is about antitone Boolean functions. Given the maximal faces  (=facets) of  $\SC$, our main algorithm, called Facets-To-Faces,  outputs $\SC$ in a compressed format. The degree of compression of Facets-To-Faces, which is programmed in high-level Mathematica code, compares favorably to both the hardwired Mathematica command {\tt BooleanConvert}, and to the hardwired BDD's provided by Python. A novel way to calculate the face-numbers from the facets is also presented.
Both algorithms can be parallelized and are applicable (e.g.) to reliability analysis, combinatorial topology, and frequent set mining. 


\section{Introduction}

While the present article focuses on bare algorithmics,  four application areas are outlined at the end of this introduction,  in Subsection 1.3. 
 We start with a broad (1.1), and then more detailed (1.2) outline of the article. 

{\bf 1.1} An {\it abstract\footnote{The adjective 'abstract' is sometimes added to make a distincion to the simplicial complexes considered in topological combinatorics. For the sake of brevity we henceforth drop 'abstract'.} simplicial complex} (also called {\it set ideal}) based on a set $W$ is a family ${\cal S}{\cal C}$ of subsets $X \subseteq W$ (called {\it faces}) such that from $X \in {\cal S}{\cal C}$, $Y \subseteq X$, follows $Y \in {\cal S}{\cal C}$. Without further mention, in this article all structures will be {\it finite}. In particular all simplicial complexes ${\cal S}{\cal C}$ contain maximal faces, called the {\it facets} of ${\cal S}{\cal C}$.  Henceforth we mostly stick to $W = [w]:= \{1,2, \cdots, w\}$.  A face of cardinality $k$ is called a $k$-{\it face}, and the set of all $k$-faces is denoted as ${\cal S} {\cal C}[k]$.   The numbers $N_k:=|{\cal S} {\cal C}[k]|$ are the {\it face-numbers} of the simplicial complex. 
The purpose of this article is to retrieve  the following data from the facets:

$(E)$ \ an enumeration of ${\cal S}{\cal C}$;

$(E_k)$ an enumeration of ${\cal S}{\cal C}[k]$ for one arbitrary $k \in [w]$;

$(C)$ \ the cardinality $N: = |{\cal S}{\cal C}|$;

$(C_{\forall k})$ the  face-numbers $N_k$ for all $k \in [w]$.

Although our four tasks can be phrased in terms of Boolean functions, speaking of simplicial complexes is, for the most part,  more illuminating.
While task $(E)$ matches $(C)$, there is a mismatch between $(E_k)$ and $(C_{\forall k})$. Here is why: If we change $(C_{\forall k})$ to the calculation of {\it one} $N_k$, then this  (essentially) is just as hard.
 Throughout the article the simplicial complex ${\cal S}{\cal C}_1 \subseteq {\cal P}[9]$ whose facets are

(1) $F_1 = \{1,2,3,9\}, \ F_2 = \{3,5,7,9\}, \ F_3 = \{2,7,9\}, \ F_4 = \{3,6,8,9\}, \ F_5 = \{2,4,8,9\}$,

serves to illustrate our algorithms.

The theoretic complexity of at least three of the problems is well known. To witness, according to [V] it is $\#P$-hard to calculate the number of models of a Boolean function $f$ given in DNF, even if $f$ is antitone\footnote{Strictly speaking that follows by de Morgan duality since Valiant only speaks about CNF's and monotone Boolean functions. Recall that $f$ is {\it monotone} if $x\le y\Ra f(x)\le f(y)$, and {\it antitone} if $x\le y\Ra f(x)\ge f(y)$.      }. Since (C) can be modelled by such $f$ (see (3)), this  implies  the $\#P$-hardness of $(C)$ and a fortiori $(C_{\forall k})$. 
Like most (unfortunately not all) authors we take {\it enumeration}  as a synonym for generation, thus not to be confused with mere counting. It might be counter-intuitive\footnote{More 'philosophy' on this matter follows in Subsection 6.3.  } that enumerating should ever be
 {\it more} tractable than counting. Yet
$(E)$ amounts to enumerate the models of a specific DNF, and enumerating the models of any DNF works in 'benign' polynomial total time, whereas (C) is $\#P$-hard.  Perhaps the complexity of  $(E_k)$ was known before but the author could not pinpoint a reference; the matter is settled anew in Theorem 2.
Our main contributions are however on the practical side; when computational efficiency  lacks a theoretic underpinning (which is to be expected in view of Valiant's results)  it will be evidenced by numerical experiments.
The main effort will go into $(E)$ and $(E_k)$. That is because we strive for a {\it compressed} enumeration in both cases. 

{\bf 1.1.1} Compression starts with the don't-care symbol '2' (other authors write $\ast$) which say  in $(1,0,2, 0)$ signifies that both bitstrings (=01-rows) $(1,0,0,1)$ and $(1,0,1,0)$ are allowed.  This leads to 012{\it -rows}. For instance, the modelset of a (Boolean) term $T$ like $x_2\wedge\ol{x_4}\wedge\ol{x_7}$ is the 012-row $r(T)=(2,1,2,0,2,2,0)$ (assuming there are 7 Boolean variables altogether). Conversely, any 012-row $r'$ of length $w$ yields a unique term $T(r')$ with at most $w$ Boolean variables. As usual $\{0,1\}^w$ is isomorphic to the powerset  ${\cal P}[w]$ of $[w]$. Thus $r(T)$ above can be viewed as a 16-element interval (also called 'cube') of
${\cal P}[7]$, with smallest element $\{2\}$ and largest element $\{1,2,3,5,6\}$. Suppose a Boolean function $f$ has a DNF which is {\it orthogonal} in the sense that the conjunction $T_i\wedge T_j\ (i\neq j)$ of any two terms in it is insatisfiable. Then the modelset $Mod(f)\s \{0,1\}^w$ is a disjoint union of the 012-rows $r(T_i)$. Although 'orthogonal DNF' and 'exclusive sum of products (ESOP)' are often used synonymously, in the present article ESOP always refers to a representation of $Mod(f)$ as a disjoint union of 012-rows. 

Apart from '2' novel types of wildcards will be introduced. We mainly deal with 012e-rows but in the last Section  glimpse at 012men-rows like $(e, n, m, 2, n, e, m, 0, m, e)$ (Table 11). Here $e..e, n..n, m..m$ respectively mean: at least one 1 here (so $(e,e):=\{(0,1),(1,0),(1,1)\}$ ); at least one 0 here; at least one 1 {\it and} one  0 here.

{\bf 1.2} Here comes the Section break-up. Section 2 deals with (C). After dispensing with inclusion-exclusion we turn to so-called Binary Decision Diagrams (BDD's), that will accompany us throughout the article. We use $\SC_1$ to illustrate the basic structure of BDD's and how they solve (C). The third method handling (C) applies the e-algorithm of [W2], whose main features are quickly reviewed. Section 3 is dedicated to $(C_{\forall k})$. Inclusion-exclusion can still be used but remains awfully slow. As to BDD's, an elegant method of Knuth is mentioned. The third method (with the prosaic name e+rp+sub) again exploits  the $e$-algorithm and adds another gadget. The core  Section 4 deals with (E).
We start with two naive (yet intriguing) methods solving (E).  Then come binary decision diagrams, which offer some compression via 012-rows.
Our Facets-To-Faces algorithm does better by employing 012e-rows and, as opposed to BDD's, it has a theoretic backbone (Theorem 1). Connections to combinatorial topology and convex polytopes are pointed out. The numerical experiments in Section 5 show that Facets-To-Faces always compresses better than the Mathematica command {\tt BooleanConvert}, and way better than BDD's. Timewise Facets-To-Faces keeps at bay BDD's but yields to {\tt BooleanConvert} if instead of few large facets there are many small facets. The fact that Facets-To-Faces is programmed in high-level Mathematica, whereas {\tt BooleanConvert} is 'hardwired', admittedly does not fully account for  this. But then again, it matters little since Facets-To-Faces is easy to parallelize. Section 6 offers two algorithms for $(E_k)$.  While polynomial total time can be proven for one, the other performs better in practice (due to compression).

The last two Sections can be viewed as 'side-shows'.
Section 7  investigates what happens when instead of the facets the minimal non-faces of a simplicial complex are given. The four problems $(E),(E_k),(C),(C_{\forall k})$ can then be handled in a more or less dual fashion.
 Section 8 harks back to Section 4 and makes first strides to lift Facets-To-Faces  from antitone DNF's (=simplicial complexes) to arbitrary DNF's.

{\bf 1.3} Here come four areas of application; the latter two are currently of a more tentative nature.

First  {\bf Reliability Analysis}. In this domain the usual name for 'simplicial complex' is 'coherent system' (or 'independence system'). The {\it reliability polynomial} of a coherent system
$\SC\s {\cal P}[w]$ is defined as  $RP(z):=\sum_{k=0}^w N_k z^k (1-z)^{w-k}$
where the $N_k$'s are the face-numbers of $\SC$ (see above). In several areas of engineering  (e.g. network analysis or stack filters for nonlinear signal processing) 
it is important to calculate $RP(z)$ fast, and many methods have been proposed in the last six decades. Some of them (like our e+rp+sub) target the face-numbers. In another vein, a partitioning of $\SC$ into few intervals (=012-rows) would yield $RP(z)$ immediately. Such a partitioning was found in [BN] for {\it matroid-complexes}, i.e simplicial complexes consisting of all independent sets of a matroid. Our Facets-To-Faces succeeds for {\it every} simplicial complex and  uses more powerful 012e-rows.

This leads  to {\bf Combinatorial Topology}. Namely, the number of 012-rows used in [BN] is as small as it can possibly be;  it equals the number of bases of the matroid. Generally a simplicial complex with $h$ facets is called {\it partitionable} if it can be represented as a disjoint union of $h$ many 012-rows. This is a popular concept in combinatorial topology. Many deep connections to other concepts have been established. For instance: $matroid$-$complex\Ra shellable\Ra Cohen$-$Macaulay$ and $shellable\Ra partitionable$. The long conjectured  implication $Cohen$-$Macaulay \Ra partitionable$ was falsified in [DKM]. A few ideas  on how Facets-To-Faces and e+rp+sub may touch upon these matters follow in Section 4.4.

 Third, consider the classic {\bf Inclusion-Exclusion} formula with its exponentially many summands. It is vexing  that many summands are often zero, but pleasant that the nonzero summands match a simplicial complex (aka 'nerve'). Isolation and compression of the nerve speed up classic inclusion-exclusion. See arXiv:1309.6927.

Last but hardly least, a  prominent area of data mining is {\bf Frequent Set Mining}. Specifically,  Facets-To-Faces  can compress all frequent sets from a knowledge of either the maximal frequent sets (i.e. the facets), or the minimal infrequent sets (Sec.7).  Many algorithms (e.g. the {\it A priori} method, listed in [WK]) have been proposed for these problems; all  proceeding  one-by-one. See arXiv1910.14508, which also discusses how to get the maximal frequent sets in the first place.

\section{Calculating the cardinality of ${\cal S}{\cal C}$ from its facets}

After inclusion-exclusions (2.1) and BDD's (2.2), a novel method to solve (C) is introduced 2.3.

{\bf 2.1}  Consider the simplicial complex $\SC_1$ whose $h=5$ facets are listed in (1). Using inclusion-exclusion one finds
 
 $(2)\quad |\SC_1|=|F_1|+\cdots+|F_5|-|F_1\cap F_2|-\cdots -|F_4\cap F_5|+\cdots-|F_1\cap \cdots\cap F_5|=52.$
 
 Having complexity $O(2^hw)$, this method is only efficient for small $h$, but for such $h$ has the advantage that the cardinalities of the faces $F_i$ hardly matter, as opposed to competing methods.

{\bf 2.2} Another established method uses Binary Decision Diagrams (BDD's); we recommend $\hfill$ [K,Sec.7.1.4] as a general reference. To warm up with Boolean functions and to survey the essentials of BDD's, consider this (antitone) Boolean function:

 $(3)\quad \psi_1(x_1, \cdots, x_9): = (\ol{x}_4 \wedge \ol{x}_5 \wedge \ol{x}_6 \wedge \ol{x}_7 \wedge \ol{x}_8) \vee
(\ol{x}_1 \wedge \ol{x}_2 \wedge \ol{x}_4 \wedge \ol{x}_6 \wedge \ol{x}_8)\\ \hspace*{1cm}\vee 
(\ol{x}_1 \wedge \ol{x}_3 \wedge \ol{x}_4 \wedge \ol{x}_5 \wedge \ol{x}_6\wedge \ol{x}_8 ) \vee
(\ol{x}_1 \wedge \ol{x}_2 \wedge \ol{x}_4 \wedge \ol{x}_5 \wedge \ol{x}_7) \vee 
  (\ol{x}_1 \wedge \ol{x}_3 \wedge \ol{x}_5 \wedge \ol{x}_6 \wedge \ol{x}_7).$

The {\it models} $x$ of $\psi_1$ (i.e. the bitstrings $x\in \{0,1\}^9$ with $\psi_1(x)=1$)
 match the faces of $\SC_1$.
 For instance $\{2,8\}\in\SC_1$ since $\{2,8\}\s F_5$. Accordingly 
 
 $\psi_1(0,1,0,0,0,0,0,1,0)=(1\wedge 1\wedge 1\wedge 1\wedge 0)\vee (1\wedge 0\wedge 1\wedge 1\wedge 0)\vee  (1\wedge 1\wedge 1\wedge 1\wedge 1\wedge 0)\vee\\ \hspace*{4.2cm}   (1\wedge 0\wedge 1\wedge 1\wedge 1)\vee (1\wedge 1\wedge 1\wedge 1\wedge 1)\\ \hspace*{3.6cm}
 =0\vee 0\vee 0\vee 0\vee 1=1 .$
  
   On the other hand, $\{1,2,8\}\not\in\SC_1$ and accordingly
$\psi_1(1,1,0,0,0,0,0,1,0)=0\vee 0\vee 0\vee 0\vee 0=0.$

Whether or not a bitstring $x$ is a model of a Boolean function can  (excluding trivial cases) be decided faster by feeding $x$ to the BDD than by evaluating a potentially large Boolean formula. The BDD of $\psi_1$ is rendered in Figure 1. If $x=(x_1,..,x_9)=(0,0,1,0,0,0,0,0,1)$, then $x_1=x[1]=0$ tells us that at the top node (={\it root}) $x[1]$ of the BDD we must take the dashed branch (it being labelled by 0). It leads us to one of the two {\it sons} of $x[1]$, i.e. the one labelled $x[2]$. Since $x_2=0$, the dashed path leads us to a node labelled $x[3]$. Since $x_3=1$ we now take the solid path (it being labelled by 1), which brings us
to the rightmost node labelled $x[4]$. Because $x_4=x_5=x_6=0$, three dashed paths bring us to a node labelled $x[7]$.
Because $x_7=0$, the dashed path brings us to the {\it leaf} 1 (distinguished from ordinary nodes by a square frame). By construction of the BDD that signifies $\psi_1(x)=1$. (Notice that the values of $x_8,x_9$ were irrelevant.) One checks that indeed $\{3,9\}\in\SC_1$.  If the value of $x_4$ had been 1 instead of 0,
then we would have reached the leaf 0 (with square frame) at once, indicating that $\psi_1(0,0,1,1,0,0,0,0,1)=0$.

{\bf 2.2.1} BDD's  allow to determine the number of models fast. For this purpose we assign in a recursive manner a probability to each node. One starts by assigning probability 0 to the leaf 0, and probability 1 to the leaf 1. Working one's way from bottom to top,  if $\alpha$ has sons $\beta,\gamma$ with probabilities $p_\beta,p_\gamma$, assign to it probability $p_\alpha:={\frac{1}{2}}p_\beta+{\frac{1}{2}}p_\gamma$.
For $\psi_1$ in the end the root gets probability $\frac{13}{128}$. Since the total number of length 9 bitstrings is $2^9$, a moment's thought shows that the cardinality of the model set $Mod(\psi_1):=\{x\in\{0,1\}^9:\ \psi_1(x)=1\}$ is
${\frac{13}{128}}\cdot 2^9=52$, which matches (2). The cost of calculating (C) this way is linear in the {\it size} of the BDD (=number of its nodes).

\includegraphics[scale=0.4]{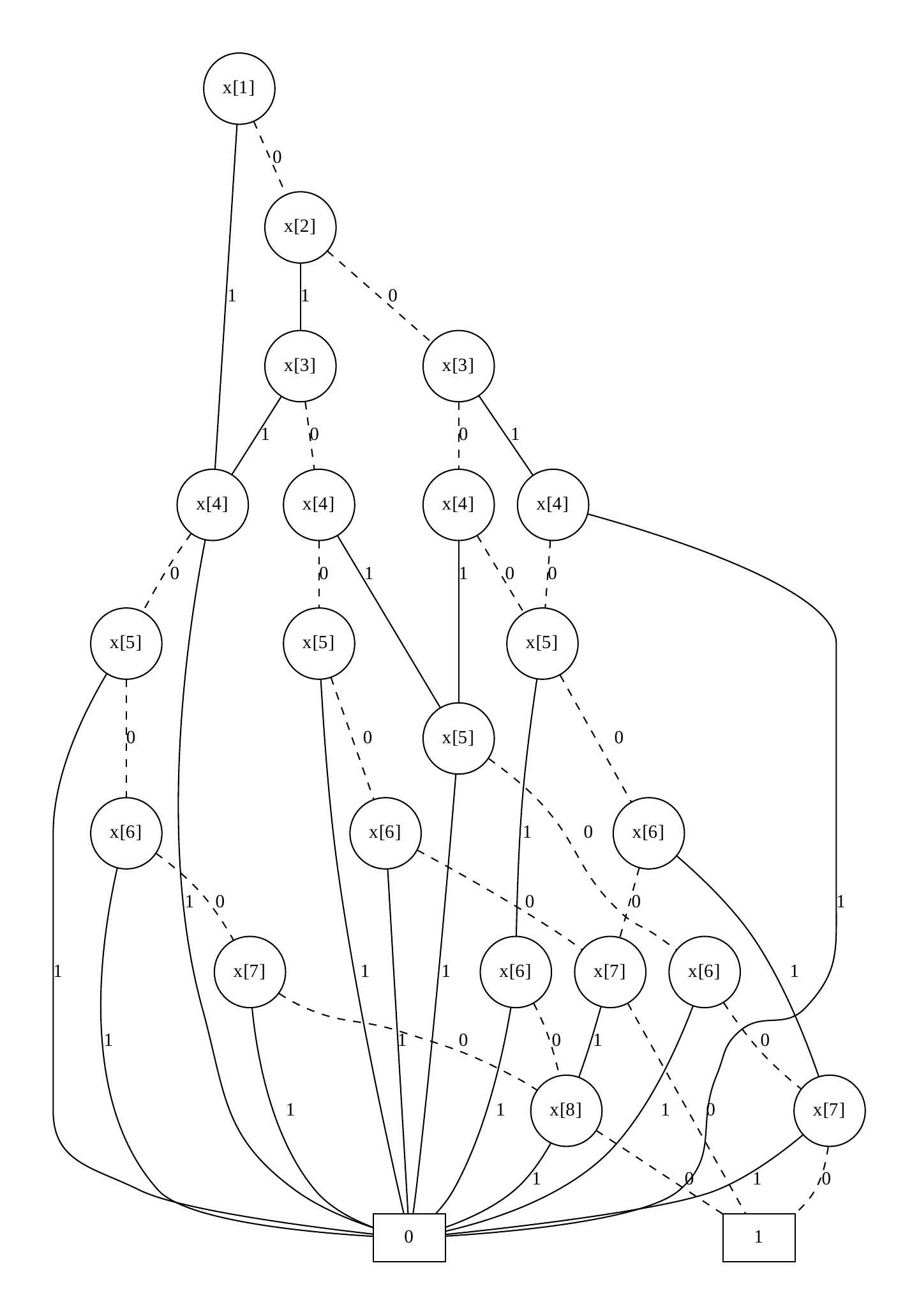} 

{\sl Figure 1: One (of many) BDD of $\psi_1$ in (3)}
 
{\bf 2.3} The third way to settle (C) is based on a certain e-algorithm, which in turn is based on 012e-rows. Extending the concept of a 012-row (Introduction), by definition a {\it 012e-row}  contains one or more wildcards of type $(e,e,..,e)$, each one of  which demanding 'at least one 1 here'. Thus the 012e-row $(e,0,1,e,2)$ is the set of bitstrings
$(e,0,1,e,{\bf 0})\cup(e,0,1,e,{\bf 1})$, where e.g. $(e,0,1,e,0):=\{({\bf 0},0,1,{\bf 1},0),({\bf 1},0,1,{\bf 0},0),
({\bf 1},0,1,{\bf 1},0)\}$. If several wildcards occur, they are distinguished by subscripts. Calculating the number of bitstrings contained in a 012e-row is easy, say $|(1,e_1,0,e_1,e_1,2,e_2,e_1,2,e_2)|=2^2\cdot(2^4-1)\cdot(2^2-1)=180.$

Recall that a {\it transversal} of a hypergraph (=set system) ${\cal H}\s {\cal P}(W)$ is a subset $X\s W$ such that $X\cap H\neq\emptyset$ for all $H\in {\cal H}$. Let $\T({\cal H})$ be the set of all transversals. The {\it(transversal) e-algorithm}, fully described in [W2], represents $\T({\cal H})$ as a disjoint union of $R$ many 012e-rows in polynomial total time $O(Rh^2w^2)$.

{\bf 2.3.1 } Consider now any simplicial complex ${\cal S}{\cal C} \subseteq {\cal P}(W)$ with facets $F_1, F_2,...$ and so on. Putting $Z^c : = W\setminus Z$ for any $Z \subseteq W$ it holds for all $X\s W$ that

(4) \quad $X \not\in {\cal S}{\cal C} \ \Leftrightarrow \ (\forall i) X \not\subseteq F_i \ \Leftrightarrow \ (\forall i) (X \cap F^c_i \neq \emptyset)$.
 
To fix ideas, take $\SC={\cal SC}_1$, whose five facets $F_i$ are listed in (1). If we apply 
the  $e$-algorithm  to ${\cal  H}=\{H_1,\ldots,H_5\}: = \{F^c_1, \cdots, F^c_5\}$ then  it outputs  $\T({\cal H})$ as a disjoint union of seven $012e$-rows:

\begin{tabular}{l|c|c|c|c|c|c|c|c|c|} 
& 1 & 2 & 3 & 4 & 5 & 6 & 7 & 8 & 9\\ \hline
&  &   &   &    &   &   &   &  & \\ \hline
$r_1'=$ & $e$ & 2 & $e$ & 1 & $e$ & $e$ & $e$ & 2 &2 \\ \hline
$r_2'=$ & 0 & 1 & 2 & 0 & 0 & 1 & 0 & 2 & 2\\ \hline
$r_3'=$ & $e_1$ & 2 & 2 & 0 & $e_1$ & $e_2$ & $e_1$ & $e_2$ & 2\\ \hline
$r_4'=$ & 0 & 1 & 1 & 0 & 0 & 0 & 0 & 1 & 2\\ \hline
$r_5'=$ & 0 & 1 & 2 &0 & 1 & 0 & 2 & 0 & 2 \\ \hline
$r_6'=$ & 1 & 2 & 2 & 0 & $e$ & 0 & $e$ & 0 & 2\\ \hline
$r_7'=$ & 0 & 1 & 1 & 0  &0 &  0 & 1 & 0 & 2\\ \hline  \end{tabular} 

{\sl Table 1: Compressing  ${\cal P}(W) \setminus {\cal SC}_1$ with the transversal $e$-algorithm}

According to (4), $\T({\cal H})$ coincides with the set filter ${\cal S}{\cal F}_1 : = {\cal P}[9] \setminus {\cal SC}_1$. It follows that 

(5) \quad $|{\cal SC}_1| = 2^9 - |{\cal S}{\cal F}_1|  =2^9 -|r'_1| - |r'_2| - \cdots -  |r'_7|$

\hspace*{1.3cm} $= 2^9 - 2^3 (2^5-1) - 2^3- 2^3 \cdot 7 \cdot 3 -2 -2^3 - 2^3 \cdot 3 - 2 = 512 - 460 = 52$,

which matches the number obtained in 2.1 and 2.2. As will be seen in 3.3.1, inclusion-exclusion stands no chance against the method of 2.3. The bottleneck in 2.2 is the calculation of the BDD {\it itself}.
That's because the expected\footnote{To be fair, in many scenarios the occuring Boolean functions do not represent a random sample of all $2^{2^w}$ many Boolean functions, and the BDD-size can be moderate then.   } size of the BDD of a Boolean function $\{0,1\}^w\to \{0,1\}$ is $2^w/w$, and hence the calculation of BDD's cannot be done in polynomial total time; the numerical experiments in Section 5 will speak the same language.

\section{Calculating the face-numbers of $\SC$ from its facets}

Here we settle $(C_{\forall k})$ by refining the three methods of Section 2.

{\bf 3.1} Generalizing (2) the principle of inclusion-exclusion als applies to calculate the face-numbers $N_k$. Thus for $\SC=\SC_1$ we find

$(6)\quad N_k={|F_1| \choose k}+\cdots +{|F_5| \choose k}-{|F_1\cap F_2| \choose k}-\cdots+{|F_4\cap F_5| \choose k}+\cdots+{|F_1\cap\cdots\cap F_5| \choose k}.$

For say $k=3$ this gives

$(7)\quad N_3={4\choose 3}+{4\choose 3}+{3\choose 3}+{4\choose 3}+{4\choose 3}-0-\cdots+0=17.$

{\bf 3.2 } While calculating the number of models of a Boolean function from its\footnote{We mention in passing that 'its' is unprecise.  A Boolean function has a  unique BDD only once a linear ordering of the Boolean variables has been fixed. The BDD in Figure 1 is based on the (popular default) ordering $x_1,...,x_w$.} BDD  is well known, using BDD's to calculate the number of models of fixed Hamming weight (here: faces of fixed cardinality) is less known.  Somewhat streamlining the account of Knuth [K,p.260,Exercise 25], details can be found in the preprint arXiv.1703.08511v5. Suffice it to say that the sought numbers $N_k$ fall out as the coefficients of a certain polynomial that is calculated recursively by processing the BDD bottom-up, in much the same way as in 2.2.

{\bf 3.3 } One ingredient of the third method for $(C_{\forall k})$ will also exploit coefficients of  polynomials, but they are  different from Knuth's polynomials. The main ingredient is, as in 2.3, the $e$-algorithm. Consider thus a generic $012e$-row 

(8) \quad $r\quad =\quad (\underbrace{0, \cdots, 0}_\alpha , \underbrace{1, \cdots, 1}_\beta , \underbrace{2, \cdots, 2}_\gamma , \underbrace{e_1, \cdots, e_1}_{\varepsilon_1} , \cdots , \underbrace{e_t, \cdots, e_t}_{\varepsilon_t})$

It is easy to see that the number Card$(r,k)$ of $k$-element sets in $r$ equals the coefficient of $x^k$ in the {\it row-polynomial}

(9) \quad $p(x) : = x^\beta \cdot (1+x)^\gamma \cdot [(1+x)^{\varepsilon_1} -1] \cdot [(1+x)^{\varepsilon_2} -1 ] \cdots [(1+x)^{\varepsilon_t} -1].$

Details on the complexity of calculating these coefficients can be found in [W2, Theorem 1]. Here we simply apply the Mathematica command {\tt Expand} to the polynomial induced by $r=r_3'$ in Table 1 and obtain

(10) \quad $p(x) = (1+x)^3 (3x + 3x^2 + x^3)(2x +x^2) = 6x^2 + 27x^3 + 50x^4 + 49x^5 + 27x^6 + 8x^7 + x^8$.

 Thus e.g. Card$(r'_3, 3) = 27$. Recall from 2.3.1 that $H_i:=F_i^c$. Let $\tau_k$ be the number of $k$-element transversals of $\{H_1, \cdots, H_5\}$, i.e. the number of $k$-element sets of ${\cal S}{\cal F}_1$. By the above, all numbers $\tau_k$ are readily calculated as

(11) \quad $\tau_k = \, \mbox{Card}(r'_1, k) + \mbox{Card}(r'_2, k) + \cdots + \mbox{Card}(r'_7, k).$ 

 Hence the face-numbers $N_k$ of ${\cal S}{\cal C}_1$ (or any simplicial complex) can be calculated with this 'subtraction trick':

(12) \quad $N_k = {9 \choose k} - \tau_k \quad (1 \leq k \leq 9).$

For instance $N_3 = {9 \choose 3} - \tau_3 = 84 - (25 + 3 + 27 + 1 + 3 + 7 + 1) = 17$, which matches (7).
In view of the \#P-hardness of $(C_{\forall k})$ and the costly calculation of BDD's we consider our threefold approach  

$\hbox{\sl e-algorithm}\ + \ \hbox{\sl row-polynomials}\ +\ \hbox{\sl subtraction trick (e+rp+sub)}$ 

 a nice way to get the face numbers from the facets. The e-algorithm is easy to parallelize (by the same reason as in [W4, sec.6.5]), and therefore also e+rp+sub. In contrast, the calculation of a BDD from a Boolean formula can hardly be parallelized.
 
{\bf 3.3.1} In a previous version of the present article (arXiv:1302.1039v4)  e+rp+sub was pitted against inclusion-exclusion on random simplicial complexes of type $(w,h,fs)$, i.e. the $h$ facets $F_i\s{\cal P}[w]$ all had facet-size $|F_i|=fs$. Predictably inclusion-exclusion  took time almost proportional to $2^h$;
 thus $(w,h,fs)=(1200,15,200)$ took 164 seconds and $(1200,20,200)$ needed to $5265  \approx 2^5\cdot 164$ seconds. In contrast, e+rp+sub took 1896 sec for the latter, and handled $(1200,40,200)$ (which triggered about 2 billion 012e-rows) in 64606 seconds . The corresponding time for inclusion-exclusion measures in centuries.

\section{The Facets-To-Faces algorithm}

Here we tackle the main task (E), i.e. given the facets, enumerate (preferably compressed)  all faces! Section 4.1 describes two naive algorithms. The first is everybody's first temptation, but  outputs the faces one-by-one. Although the second has the potential for compression (using 012-rows), it nevertheless can be inferior.  Knowing a BDD of  $\SC$ one can use 012-rows more efficiently for compression. Section 4.3 introduces the novel Facets-To-Faces algorithm which displays $\SC$ as a disjoint union of more powerful 012e-rows. Section 4.4 relates Facets-To-Faces to facets and faces  of {\it topological} simplicial complexes and convex polytopes. 

{\bf 4.1 } We put in front some definitions for 4.1.2. If $\SC\s {\cal P}[w]$, call a  012-row $r$  of length $w$ {\it feasible} if $r\cap{\cal S}{\cal C}\not=\emptyset$ (which amounts to $ones(r)\subseteq F_i$ for some $i$). Further call $r$ {\it final} if  $r\subseteq{\cal S}{\cal C}$ (which amounts to $ones(r)\cup twos(r)\subseteq F_i$ for some $i$).

{\bf 4.1.1 } The {\it First Naive Algorithm (FNA)} for $(E)$ enumerates  ${\cal S}{\cal C}_1$ simply as
$\SC_1={\cal P}(F_1) \cup \cdots \cup {\cal P}(F_5)$. As to 'simply', trouble is that multiple occurencies of faces (such as $\{2,9\}\in {\cal P}(F_1)\cap {\cal P}(F_3)\cap {\cal P}(F_5)$ need to be pruned. Specifically, by induction suppose that for any $\SC$ we have obtained ${\cal P}(F_1)\cup\cdots\cup{\cal P}(F_k)=\{A_1,...,A_t\}$ such that $A_i\neq A_j$ for $i\neq j$. Then only the members of ${\cal P}(F_{k+1})$ distinct from all $A_i$'s are added to the list. Here comes a confession: This is how FNA {\it de iure} must be programmed. De facto we exploited two shortcuts provided by Mathematica.
First, enumerating a powerset (such as ${\cal P}(F_{k+1})$ ) is more subtle than it looks; see [K,sec. 7.2.1.1]. We circumvented that issue with the Mathematica command {\tt Subsets}, which behaves as follows: {\tt Subsets}$[\{3,7\}]$ directly outputs
$\{\{\},\{3\},\{7\},\{3,7\}\}$. Second, the command {\tt Union} automatically prunes multiple occurencies (and orders the output); thus {\tt Union}$[\{4,11,9,2\},\{2,11,7\}]$ outputs $\{2,4,7,9,11\}$. Therefore, if ${\cal P}(F_1)\cup\cdots\cup{\cal P}(F_k)=\{A_1,...,A_t\}$ has been computed, we get a pruned listing of ${\cal P}(F_1)\cup\cdots\cup{\cal P}(F_{k+1})$ with {\tt Union}$[\{A_1,...,A_t\},{\tt Subsets}[F_{k+1}]]$.

{\bf 4.1.2} The  {\it Second Naive Algorithm (SNA)} uses variable-wise branching (a better name being {\it pivotal decomposition}, as argued in [W4, sec. 2.5]).
Initially our Last-In-First-Out (LIFO) stack only contains the feasible row $(2,2,\ldots,2)$. Generally always the top 012-row $r$ of the LIFO-stack is picked. The "first" occuring digit 2 (with respect to a fixed ordering of the index set $\{1,2,...,w\}$) is turned to 0 and 1 respectively. This yields 012-rows $r_0$ and $r_1$. By induction $r$ was feasible. Since subsets of faces are faces, it follows that $r_0$ is feasible, but not necessarily $r_1$. These one or two feasible 012-rows replace $r$ on the LIFO stack
(except that final rows go to an initially empty 'final stack').
 As soon as the LIFO stack is empty, the union of the 012-rows in the final stack is disjoint and equals ${\cal S}{\cal C}$.  (Theorem 2 fine-tunes the above in a more sophisticated setting.)

{\bf 4.1.3 } Here comes an experimental comparison of the two naive algorithms. For various  random instances $(w,h,fs)$ (see 3.3.1) we recorded the times (rounded to full seconds) $T_1,\ T_2$ needed for FNA and SNA respectively to enumerate the ensuing simplicial complex ${\cal S}{\cal C}$. The number $R_2$ of final 012-rows produced by SNA is recorded as well. In contrast, FNA offers no compression but, recall, its advantage is that
all faces contained in a facet $F_i$, are 'instantaneously' produced by  {\tt Subsets[$F_i$]}. This advantage wins out in the $(20,100,10)$ instance. The $(20,10,18)$-instance lets SNA catch up because it compresses on average roughly 16 faces per 012-row, whereas FNA outputs faces one-by-one and invests considerable time (despite the hardwired {\tt Union} command!) to prune duplicated faces. These two trends increase in the $(20,1000,16)$-instance to the extent that SNA is more than twice as fast as FNA. The tables are turning again in the extrapolated $(24,1000,16)$-instance because the compression of SNA is just too poor (about 2 faces per 012-row). Finally, the extrapolated $(50,100,20)$-instance with its 109'437'738 faces\footnote{Several of the methods dscussed in Section 5 can get this number very fast} provides a Pyrrhus victory for SNA:
The FNA ran out of memory while executing {\tt Union}$[\{...\},{\tt Subsets}[F_{19}]]$.

\begin{tabular}{l|c|c|c|c|c|c|c|} 
	& w & h & fs & $|{\cal S}{\cal C}|$ & $T_1$ & $T_2$ & $R_2$ \\ \hline
	&  &   &   &   &   &    \\ \hline
	$ $ & 20  & 100 & 10 & 51'489 & 1 & 63& 26'624 \\ \hline
	$ $ & 20 & 10 & 18 & 939'136 & 3 & 8 & 57'314 \\ \hline
	$ $ & 20 & 1000 & 16 & 1'035'899 & 408 & 161 & 23'567 \\ \hline
	$ $ & 24 & 1000 & 16 & 8'529'330 & 2'327& 22 hrs &  4'000'000 \\ \hline
	$ $ & 50 & 100 & 20 & 109'437'738 & -- & 131 hrs & 76'000'000\\ \hline
\end{tabular}

{\sl Table 2: Comparison of the two naive algorithms }

{\bf 4.1.4 } For SNA the number of final 012-rows which are proper (i.e. not 01-rows) heavily depends on the particular ordering of the index set $\{1,2,..,9\}$. For instance, using the natural ordering $1,2,..,9$ the SNA represents our 52-element example ${\cal S}{\cal C}_1$ as a disjoint union of 19 rows. The minimum (=13) and maximum (=44) number of final 012-rows are obtained (e.g.) for the orderings $1,2,4,5,7,6,8,3,9$ and $1,2,3,4,5,8,9,7,6$ respectively.

{\bf 4.2 } Figure 1 shows the BDD of the Boolean function $\psi_1$ of (3). Recall from 2.2 that $\psi_1(x)=1$ for $x=(0,0,1,0,0,0,0,0,1)$, and so feeding $x$ to the BDD traced a path from the root $x[1]$ to the 1-leaf. The fact that the values $x_8,x_9$ were irrelevant for reaching the 1-leaf shows that $(0,0,1,0,0,0,0,2,2)\s Mod(\psi_1)$. Generally each path from $x[1]$ to the 1-leaf yields a 012-row contained in $Mod(\psi_1)$, and distinct paths induce disjoint 012-rows (why?). Therefore, if there are $t$ such paths, then $Mod(\psi_1)$ can be written as union of $t$ disjoint 012-rows. 
We call this the {\it BDD-induced ESOP} of $\psi_1$ (see 1.1.1). How to find these $t$ paths efficiently?

\begin{tabular}{l|c|c|c|c|c|c|c|c|c|} 
	& $x[1]$ & $x[2]$ & $x[3]$ & $x[4]$ & $x[5]$ & $x[6]$ & $x[7]$ & $x[8]$ & $x[9]$\\ \hline
	&   &   &   &   &  &  &   &  & \\ \hline
	$\beta_1=$ & {\bf 1}  & 2 & 2 & 0 & 0 & 0& 0 & 0 & 2\\ \hline
	$\beta_0= $ & {\bf 0} & ? & ? & ? & ? & ? & ?& ? & 2\\ \hline
		&   &   &   &   &  &  &   &  & \\ \hline
	$\beta_1=$ & 1  & 2 & 2 & 0 & 0 & 0& 0 & 0 & 2\\ \hline
	$\beta_{0,0}= $ & 0 & {\bf 0} & ? & ? & ? & ? & ?& ? & 2\\ \hline	
	$\beta_{0,1}= $ & 0 & {\bf 1} & ? & ? & ? & ? & ?& ? & 2\\ \hline	
		&   &   &   &   &  &  &   &  & \\ \hline 
			
$\beta_1=$ & 1  & 2 & 2 & 0 & 0 & 0& 0 & 0 & 2\\ \hline
\hline
$\beta_{0,0,1}=$ & 0 & {\bf 0} & 0 & 0 & 0 & 0 & 0& 2 &     2\\ \hline
$ \beta_{0,0,2}= $ & 0 & {\bf 0} & 0 & 0 & 0 & 0 & 1& 0 &     2\\ \hline
$ \beta_{0,0,3}= $ & 0 & {\bf 0} & 0 & 0 & 0 & 1 & 0& 2 &     2\\ \hline
$ \beta_{0,0,4}= $ & 0 & {\bf 0} & 0 & 0 & 1 & 0 & 2& 0 &    2\\ \hline
$ \beta_{0,0,5}= $ & 0 & {\bf 0} & 0 & 1 & 0 & 0 & 0& 2 &    2\\ \hline
$ \beta_{0,0,6}= $ & 0 & {\bf 0} & 1 & 0 & 0 & 0 & 0& 2 &     2\\ \hline
$ \beta_{0,0,7}= $ & 0 & {\bf 0} & 1 & 0 & 0 & 0 & 1& 0 &    2\\ \hline
$ \beta_{0,0,8}= $ & 0 & {\bf 0} & 1 & 0 & 0 & 1 & 0& 2 &    2\\ \hline
$ \beta_{0,0,9}= $ & 0 & {\bf 0} & 1 & 0 & 1 & 0 & 2& 0 &    2\\ \hline
\hline
	$\beta_{0,1,1}= $ & 0 & {\bf 1} & 0 & 0 & 0 & 0 & 0& 2 &      2\\ \hline
$\beta_{0,1,2}= $ & 0 & {\bf 1} & 0 & 0 & 0 & 0 & 1& 0 &         2\\ \hline
$\beta_{0,1,3}= $ & 0 & {\bf 1} & 0 & 1 & 0 & 0 & 0& 2 &        2\\ \hline
$\beta_{0,1,4}= $ & 0 & {\bf 1} & 1 & 0 & 0 & 0 & 0& 0 &          2\\ \hline
\end{tabular} 

{\sl Table 3: The ESOP induced by the BDD of $\psi_1$}

A look at Figure 1 shows that the only $\psi_1$-models $x$ with $x_1=1$ are the ones in the 012-row $\beta_1$ of Table 3.
All other $\psi_1$-models must fit the pattern of $\beta_0$. One can get rid of the first '?' in $\beta_0$ by splitting $\beta_0$ as $\beta_{0,0}\uplus\beta_{0,1}$. However, continuing in this manner can create many dead-end paths. It is better to embrace a bottom-up approach akin to  2.2.1. This would show that $\beta_{0,0}=\beta_{0,0,1}\uplus\cdots\uplus\beta_{0,0,9}$ and $\beta_{0,1}=\beta_{0,1,1}\uplus\cdots\uplus\beta_{0,1,4}$.
Hence $t=1+9+4=14$. Adding up the cardinalities of the fourteen rows $\beta_1,\beta_{0,0,1},\ldots,\beta_{0,1,4}$ yields $8+4+\cdots+2=52$, as was to be expected.

{\bf 4.2.1} The mere {\it number} $R_{BDD}$ of 012-rows in a BDD-induced ESOP can be predicted without having to calculate the ESOP. Namely, proceeding bottom-up, assign integers (instead of probabilities as in 2.2.1) to the  BDD-nodes as follows. The 0-leaf and 1-leaf receive 0 and 1 respectively.  If node $\alpha$ has sons $\beta,\gamma$ with assigned integers $i_\beta,i_\gamma$, assign to it  $i_\alpha:=i_\beta+i_\gamma$. The last number $i_{root}$ equals $R_{BDD}$. The reader is invited to verify that for the BDD in Figure 1 this procedure indeed yields $R_{BDD}=14$.

{\bf 4.3} We now embark on the third method for solving (E), it being the core of our article. Suppose ${\cal S}{\cal C}$ has facets $F_1$ to $F_h$, and by induction we have obtained for some $t \in [h-1]$ a representation

(13) \quad ${\cal P}(F_1) \cup \cdots \cup {\cal P}(F_t) = \rho_1 \uplus \rho_2 \uplus \cdots \uplus \rho_s$

with $012e$-rows $\rho_i$. If $r$ is the $02$-row matching ${\cal P}(F_{t+1})$ then evidently

(14) \quad ${\cal P}(F_1) \cup \cdots \cup {\cal P}(F_{t+1}) = (\rho_1 \setminus r) \uplus (\rho_2 \setminus r) \uplus \cdots \uplus (\rho_s\setminus r) \uplus r$,

and so the key problem is this: For a given $012e$-row $\rho\ (=\rho_i)$ and $02$-row $r$ {\it recompress} the set  difference $\rho \setminus r$ as disjoint union of $012e$-rows. Let us do away with the two extreme cases first. First, $\rho \setminus r = \rho$ iff $\rho \cap r = \emptyset$ thus iff either a 1 or $e$-wildcard of $\rho$ falls into zeros$(r)$. Second, $\rho \setminus r = \emptyset$ iff $\rho \subseteq r$, thus iff zeros$(r) \subseteq \mbox{zeros}(\rho)$. For instance $(e,e,1,2,0,0) \setminus (2,2,2,2,2,0)= \emptyset$.

\begin{tabular}{l|c|c|c|c|c|c|c|c|c|c|c|c|c|}
& 1 & 2& 3 & 4 & 5 & 6 & 7 & 8 & 9 & 10 & 11 & 12 & 13 \\ \hline
&   &  &   &   &   &    &   &   &  &   &     &   & \\ \hline
$\rho =$ & $e_1$ & $e_1$ & 2 & 2 & $e_2$ & $e_3$ & $e_3$ & $e_3$ & $e_2$ & $e_2$ & $e_1$ & $1$ & 0 \\ \hline
$r=$ & 0 & 0  & 0 & 0 & 0 & 0 & 0 & 2 & 2 & 2 & 2 & 2 & 0 \\ \hline
$\rho_1'=$ & ${\bf e}_1$ & ${\bf e}_1$ & 2 & 2 & $e_2$ & $e_3$ & $e_3$ & $e_3$ & $e_2$ & $e_2$ & 2 & 1  & 0 \\ \hline 
$\rho_2'=$ & ${\bf 0}$ & ${\bf 0}$ & ${\bf e}$ & ${\bf e}$ & $e_2$ & $e_3$ & $e_3$ & $e_3$ & $e_2$ & $e_2$ & 1 & 1 & 0 \\ \hline 
$\rho_3'=$ & ${\bf 0}$ & ${\bf 0}$ & ${\bf 0}$ & ${\bf 0}$ & ${\bf 1}$ & $e_3$ & $e_3$ & $e_3$ & 2 & 2 & 1 & 1 & 0 \\ \hline 
$\rho_4'=$ & ${\bf 0}$ & ${\bf 0}$ & ${\bf 0}$ & ${\bf 0}$ & ${\bf 0}$ & ${\bf e}_3$ & ${\bf e}_3$ & $2$ & $e_2$ & $e_2$ & 1 & 1 & 0 \\ \hline 
\end{tabular}

{\sl Table 4: Recompressing the type $(012e) \setminus (02)$ set difference $\rho \setminus r$ }

In all other cases we focus on the {\it flexible} (i.e. $\neq 0,\, 1$) symbols of $\rho$,
thus for $\rho$ in Table 4 the symbols on the positions 1 to 11. The only way for $X \in \rho$ to {\it detach} itself from (the 'plebs' in)  $r$ is to employ those flexible symbols of $\rho$ that are ``above'' a $0$ of $r$, in the sense that they occupy a position which in $r$ is occupied by $0$. For the particular $\rho$ and $r$ in Table 4 a bitstring $X\in r$ detaches itself from $r$ iff ones$(X) \cap [7] \neq \emptyset$. Depending on whether the smallest element of ones$(X) \cap [7]$ belongs to $\{1,2\}, \{3,4\}, \{5\}, \{6,7\}$ (this partition is dictated by the wildcard pattern of $\rho$), the bitstring $X$ belongs to exactly one of the {\it sons} $\rho_1', \rho_2', \rho_3', \rho_4'$. 

The powersets of the five facets $F_i$ of ${\cal S}{\cal C}_1$ (see (1)) are listed as the first five $02$-rows $r_i$ in Table 5. Applying detachment repeatedly yields:

$\begin{array}{rll}
r_1 \cup r_2                            & = & (r_1 \setminus r_2) \uplus r_2 = : r_6 \uplus r_2\\
\\
(r_6 \uplus r_2) \cup r_3               &= & (r_6 \setminus r_3) \uplus (r_2 \setminus r_3) \uplus r_3 = : ( r_7 \uplus r_8) \uplus r_9 \uplus r_3\\
\\
(r_7 \uplus \cdots \uplus r_3) \cup r_4 & = & (r_7 \setminus r_4) \uplus (r_8 \setminus r_4) \uplus (r_9 \setminus r_4) \uplus (r_3 \setminus r_4) \uplus r_4\\
\\
                                        & =: & r_7 \uplus r_8 \uplus (r_{10} \uplus r_{11}) \uplus r_{12} \uplus r_4\\
\\
(r_7 \uplus \cdots \uplus r_4) \cup r_5 & = & (r_7 \setminus r_5) \uplus (r_8 \setminus r_5) \uplus (r_{10} \setminus r_5) \uplus (r_{11} \setminus r_5) \uplus (r_{12}                                               \setminus r_5) \uplus (r_4 \setminus r_5) \uplus r_5 \\
\\
                                        & = : & r_7 \uplus r_8 \uplus r_{10} \uplus r_{11} \uplus r_{13} \uplus r_{14} \uplus r_5,
\end{array}$

From Table 5 follows, as it must, that

 $$(15)\quad |r_5| + |r_7| + |r_8| + |r_{10}| +|r_{11}| + |r_{13}|  + |r_{14}| = 16 + 8 +\cdots + 12 = 52 = |{\cal S}{\cal C}_1|.$$

 We call this algorithm  {\it Facets-To-Faces}.

\begin{tabular}{r|c|c|c|c|c|c|c|c|c|l} 
& 1 & 2 & 3 & 4 & 5 & 6 & 7 & 8 & 9 & \\ \hline
&   &   &   &   &   &   &   &   &   & \\ \hline
$r_1=$ & 2 & 2 & 2 & 0 & 0 & 0 & 0 & 0 & 2 \\ \hline
$r_2=$ & 0 & 0 & 2 & 0 & 2 & 0 & 2 & 0 & 2 \\ \hline
$r_3=$ & 0 & 2 & 0 & 0 & 0 & 0 & 2 & 0 & 2 \\ \hline
$r_4=$ & 0 & 0 & 2 & 0 & 0 & 2 & 0 & 2 & 2 \\ \hline
$r_5=$ & 0 & 2 & 0 & 2 & 0 & 0 & 0 & 2 & 2 & 16 \\ \hline
$r_6=$ & $e$ & $e$ & 2 & 0 & 0 & 0 & 0 & 0 & 2 \\ \hline
$r_7=$ & 1 & 2 & 2 & 0 & 0 & 0 & 0 & 0 & 2 & 8 \\ \hline
$r_8=$ & 0 & 1 & 1 & 0 & 0 & 0 & 0 & 0 & 2 & 2 \\ \hline
$r_9=$ & 0 & 0 & $e$ & 0 & $e$ & 0 & 2 & 0 & 2 \\ \hline
$r_{10}=$ & 0 & 0 & 2 & 0 & 1 & 0 & 2 & 0 & 2 & 8\\ \hline
$r_{11}=$ & 0 & 0 & 1 & 0 & 0 & 0 & 1 & 0 & 2 & 2\\ \hline
$r_{12}=$ & 0 & $e$ & 0 & 0 & 0 & 0 & $e$ & 0 & 2 & \\ \hline
$r_{13}=$ & 0 & 2 & 0 & 0 & 0 & 0 & 1 & 0 & 2 & 4\\ \hline
$r_{14}=$ & 0 & 0 & $e$ & 0 & 0 & $e$ & 0 & 2 & 2 & 12\\ \hline
\end{tabular}

{\sl Table 5: Compressing ${\cal S}{\cal C}_1$ with  Facets-To-Faces  }

{\bf Theorem 1:} {\it Let $F_1,\ldots,F_h\subseteq [w]$ be the facets of a simplicial complex ${\cal S}{\cal C}$. Then  Facets-To-Faces   enumerates
${\cal S}{\cal C}$ as a union\footnote{In view of the 012-rows entering the definition of 'ESOP', one could call this kind of union a 'fancy ESOP' for $\SC$ or, more precisely, for its underlying antitone Boolean function $f$ (such as (3) for $\SC=\SC_1$).  } of $R$ disjoint 012e-rows in time $O(R^2w^2h)$.}

{\it Proof.} By induction assume that for some $t<h$ the decomposition (13) has been achieved. If some 012e-row $\rho_i$ is contained in ${\cal P}(F_{t+1}) \cup \cdots \cup {\cal P}(F_h)$ then neither  $\rho_i$ nor any of its sons and grandsons will survive in the long run. Thus  $\rho_i$ is a {\it dud}, i.e. causing work without benefit.  Moreover, unless  $\rho_i$ is cancelled right away, it is impossible to predict the algorithm's total time. Fortunately, letting $X=X(i)$ be the unique largest set in  $\rho_i$ (thus $X$ is obtained by setting all $2$'s and $e$'s to $1$), it holds that

$$ \rho_i\subseteq {\cal P}(F_{t+1}) \cup \cdots \cup {\cal P}(F_h)\Leftrightarrow X\in {\cal P}(F_{t+1}) \cup \cdots \cup {\cal P}(F_h)\Leftrightarrow (\exists j\in\{t+1,\dots,h\})\ X\subseteq F_j.$$

Testing for all $1\le i\le s$ whether $\exists\ j\in\{t+1,\ldots,h\}$ with $X(i)\subseteq F_j$ costs $O(s(h-t)w)$. In other words, that is the cost of pruning the righthand side of (13) from duds. What is the cost to proceed from a (pruned) representation (13) to a (not yet pruned) representation (14)? Because $\rho_i\setminus r$ has at most $w$ sons (which is clear from Table 4), and 'writing down' each son is obvious (i.e. costs $O(w)$), the asked for cost is $O((s+1)w^2)$. Hence the overall cost is

$$R\cdot\bigg( O(s(h-t)w)+O((s+1)w^2) \bigg)=R\cdot\bigg( O(Rhw)+O(Rw^2) \bigg)=O(R^2w^2h).\ \square$$

{\bf 4.3.1} Suppose Facets-To-Faces has advanced to representing ${\cal P}(F_1)\cup\cdots\cup{\cal P}(F_{s-1})$ as a disjoint union of 012e-rows. At one's digression one can then embark on {\it distributing the computation} to $t$ satellite stations. Say $t=3$ and 

${\cal P}(F_1)\cup\cdots\cup{\cal P}(F_{s-1})=(r_1^{(1)}\uplus\cdots\uplus r_\alpha^{(1)})\uplus (r_1^{(2)}\uplus\cdots\uplus r_\beta^{(2)})\uplus (r_1^{(3)}\uplus\cdots\uplus r_\gamma^{(3)}),$

where $\alpha,\beta,\gamma$ are approximately equal. Putting $r:={\cal P}(F_s)$ the control sends $r_1^{(1)},...,r_\alpha^{(1)};r$ to satellite 1, and $r_1^{(2)},...,r_\beta^{(2)};r$ to satellite 2, and $r_1^{(3)},...,r_\gamma^{(3)};r$ to satellite 3. After a while the control receives from satellite 1 some 012e-rows
$\rho_1^{(1)},...,\rho_{\alpha'}^{(1)}$ such that $(r_1^{(1)}\setminus r)\uplus\cdots\uplus (r_\alpha^{(1)}\setminus r)=
\rho_1^{(1)}\uplus\cdots\uplus\rho_{\alpha'}^{(1)}$. Satellite 2 and satellite 3 send analogous rows $\rho_1^{(2)},...,\rho_{\beta'}^{(2)}$ and $\rho_1^{(3)},...,\rho_{\gamma'}^{(3)}$.(Note that $\alpha',\beta',\gamma'$ may differ significantly in magnitude.) The control pools the received rows, {\it adds} row $r$, and divides the $\alpha'+\beta'+\gamma'+1$ rows in three approximately equal-sized parts. The three parts, each augmented by $r':={\cal P}(F_{s+1})$, are sent back to the satellites. And so forth.

{\bf 4.4.} This Subsection links the above to convex polytopes.  We begin with the  framework of $\cap$-subsemilattices ${\cal L} \subseteq {\cal P}(W)$, i.e. $X,Y\in {\cal L}\Ra X\cap Y\in {\cal L}$. If the set ${\cal M}({\cal L})$ of meet-irreducibles (or any $\cap$-generating set) is known, then ${\cal L}$ can be generated one-by-one in polynomial total time by a variety of algorithms. These algorithms e.g. are of  interest in Formal Concept Analysis [GO]. Ganter's {\it NextClosure} algorithm [GO,p.44] was the first and is still popular.

Consider now a convex polytope $P\s \R^d$. Quoting from [FR,p.192]: {\it ...the combinatorial face enumeration problem (CFEB) is to enumerate all faces of $P$ in terms of their representations without duplications.} What Fukuda and Rosta mean by the 'representation' of a face $F$ is the set of facets in which $F$ is contained. Let $W$ be the set of vertices of $P$, identify each face of $P$ with the set of vertices it contains, and let ${\cal L}\s {\cal P}(W)$ be the set of all faces. In this setting CFEP reduces to enumerating ${\cal L}$ from the set ${\cal M}({\cal L})$ of facets. (As to how the facets {\it themselves} can be found, see 4.4.2.)
In [KP], which improves upon results in [FR], and which was inspired by NextClosure, not only the individual faces but all {\it covering pairs} of faces are enumerated from the facets in polynomial total time. 

{\bf 4.4.1} A convex polytope is a {\it simplex} if {\it any} subset of (the vertex set of) any face is (the vertex set of) a face. For instance, the simplices in $\R^3$ are exactly the tetrahedrons. Gluing together simplices yields {\it (topological) simplicial complexes}\footnote{Since we are only concerned with abstract simplicial complexes (defined in 1.1) we can dispense with a formal definition of topological simplicial complexes. We mention in passing: Other than  convex polytopes, simplicial complexes $\SC$ which are not simplices have meet-irreducible faces which are not facets (which?). However this is  irrelevant since $\SC$ is already determined by its facets.}. As is to be expected, the [KP]-algorithm accelerates for simplicial complexes, yet still enumerates one-by-one. In [BM], which similarly caters for  combinatorial topologists, the individual faces are organized in a tree-structure. This supports various combinatorial operations (such as contracting edges), but  again offers no  compression. 

Enter Facets-To-Faces. Apart from the practical aspects of compression, there is a connection to an important theoretical concept. Namely, in any disjoint representation of $\SC$ by 012e-rows each facet must be the largest member in the row $r_i$ it happens to belong to. In particular, if there are $h$ facets then any disjoint representation comprises at least $h$ many 012e-rows. In combinatorial topology a simplicial complex is called {\it partitionable} if one can do with $h$ many 012-rows (yet other terminology is used). The relevance of this concept has been indicated in 1.3. Here are two veins for further research. Can the methods in arXiv:1811.11689   (which concern shellability) be adapted to find necessary or sufficient conditions for the partitionability of a random simplicial complex given by its facets? Defining $\SC$ to be {\it e-partitionable} if $\SC$ is a disjoint union of $h$ many 012{\bf e}-rows, is this notion strictly weaker than being partitionable?

{\bf 4.4.2} Notice that $\cap$-subsemilattices ${\cal L}\s {\cal P}(W)$ (e.g. arising from convex polytopes) are not easily compressed; it seems one needs an implicational base $\Sigma$ of the closure system ${\cal L}\cup\{W\}$, but calculating $\Sigma$ from ${\cal M}({\cal L})$ is usually hard [W3, sec. 3.6].  Is there nevertheless a place for compression in the context of convex polytopes $P\s \R^d$? To answer, recall the two fundamental representations of  $P$. The {\it  H-representation} views $P$ as an intersection of  $h$ closed half-spaces, each one of which given by an inequality $a_{i,1}x_1+\cdots+a_{i,d}x_d\le b_i$. The{\it  V-representation} gives the  vertex set $W\s \R^d$, viewing that $P$ is the convex hull of $W$. Much research\footnote{In particular, various methods have been proposed to get the 'combinatorial' facets (i.e the sets of incident vertices) from either the H- or the V-representation.          } has been devoted going from one kind of representation to the other. 
As to 'a place for compression', given the H-representation of  $P$, there is hope to compress the set of  interior 0,1-points (i.e. $P\cap \{0,1\}^d$) as a disjoint union of 012e-rows (work in progress).

\section{Numerical experiments}

Theorem 1 only implies $R\le | {\cal S}{\cal C}|$ (due to the disjointness of 012e-rows), yet the numerical experiments below show that often $R<< | {\cal S}{\cal C}|$. The meaning of a random instance $(w,h,fs)$ is as in 3.3.1 and 4.1.3. The number $R$ of final 012e-rows spawned by the Facets-To-Faces algorithm, and its running time $T$ in seconds are recorded.  In our implementation of Facets-To-Faces  the precaution to avoid duds (see the proof of Theorem 1) was omitted because for the instances in Table 7 its incorparation would outweigh the benefits. For instance the (50,240,20)-instance features 460631 final versus 13244 wasteful 012e-rows. In the other instances the proportion wasteful/final is even smaller. In all instances more than half of the final 012e-rows were proper, i.e. not 012-rows. In the (2000,70,192)-instance only 1157 out of 70551 many 012e-rows were improper. 

After introducing the two competitors of Facets-To-Faces (5.1), we assess the three algorithms' compression capabilities in 5.2. Then we compare  with respect to CPU time (5.3), and finally with respect to memory requirements (5.4).

{\bf 5.1}  Mathematica uses BDD's only behind the scenes, in particular for the command $\hfill$ {\tt SatisfiabilityCount} which outputs the number\footnote{Once Facets-To-Faces terminates, the cardinality $|\SC|$ is easily determined (see 5.5), and it always coincided with the number produced by  {\tt SatisfiabilityCount}. Hence Facets-To-Faces  very likely works correctly.}   of models of any Boolean function fed to it. Therefore I am grateful to Maximilian Vides, who helped me to access BDD's  via Python. Specifically, for each instance $(w,h,fs)$ the matching  antitone Boolean function was translated from Mathematica notation to Python notation, then fed to the Python command {\tt expr2bdd} which calculates a BDD, then $R_{BDD}$ was calculated as described in 4.2.1. The first competitor {\tt expr2bdd} always uses  the natural default ordering $x_1,...,x_w$ of variables. This may be part of the explanation why it is much slower than {\tt SatisfiabilityCount}. In any case, even if the BDD underlying  {\tt SatisfiabilityCount}$[f]$, undergoes minimisation,  it seems to induce an ESOP that compresses poorly\footnote{ Much research has gone into optimizing the variable ordering in order to reduce (or even minimize) the size $s(BDD)$ and whence the time to calculate it. However, there is little relation between $s(BDD)$ and $R_{BDD}$ because it is the structure rather than the sheer number $s(BDD)$ of nodes that determines $R_{BDD}$.  For instance $s(BDD)=4601>R_{BDD}$ for $(50,30,10)$ but  $s(BDD)=765'720<R_{BDD}$ for $(50,40,30)$. It seems that no research has gone into optimizing the variable ordering to make $R_{BDD}$ small. It may well be that this is fruitless since competing methods will keep on compressing $Mod(f)$ better.  Likewise  the compression achieved by Faces-To-Facets  heavily depends on the order in which the facets are processed, and so far no research in this regard exists. We conclude: Facets-To-Faces is just as disadvantaged by the default ordering of faces as {\tt expr2bdd} is by the default variable ordering.  }. Why else would Mathematica use a command which is {\it not} based on BDD's (as confirmed by the author) to calculate an ESOP of a Boolean function? This command is {\tt BooleanConvert} (option 'ESOP'), and it is the second competitor of Facets-To-Faces.

{\bf 5.2} In Table 6 the parameter $R$ is the number of 012e-rows produced by Facets-To-Faces, $R_{BC}$ is the number of 012-rows in the ESOP calculated with {\tt BooleanConvert}, and $R_{BDD}$ was defined above. One sees that for {\it all} instances $(w,h,fs)$ it holds that $R<R_{BC}<R_{BDD}$. 
 Let us  fix $w$ and $h$ and observe what happens when $fs$ increases. If say $(w,h)=(50,10)$, then $(50,10,{\bf 10})\ra (R,R_{BC},R_{BDD})=(39,200,440)$ and $(50,10,{\bf 25})\ra (285,5992,62503)$ and $(50,10,{\bf 40})\ra (429,220'432,2'389'705)$. Likewise fixing $(w,h):=(50,300)$ and taking $fs=10,17,20,26$ gives a similar picture, except that {\tt expr2bdd} couldn't finish in reasonable time.
In brief, all else being equal, all three suffer from increasing $fs$, {\tt expr2bdd} more than  {\tt BooleanConvert}, and {\tt BooleanConvert} more than Facets-To-Faces.
We leave it to the reader to draw conclusions (although the data is sparse) from fixing $w,fs$ and increasing $h$, respectively fixing $h,fs$ and increasing $w$.

{\bf 5.3} What concerns CPU-times, the state of affairs is not so clear-cut. In a nutshell, the Facets-To-Faces algorithm  dislikes many short facets, but likes few large facets.
As to few but large facets, in such situations it may not only best the time of {\tt BooleanConvert} but even {\tt SatisfiabilityCount}: It took Facets-To-Faces $T=1114$ seconds to  squeeze $10^{92}$ faces (contained in 70 facets of size 300) into a mere 707518 many 012e-rows, whereas  {\tt SatisfiabilityCount} (which we only asked to {\it count} the faces) was aborted after fourteen hours. 
When there are many small facets  (such as for $(w,h,fs)=(50,1000,10)$ ) then $T_{BC}$ is smaller than $T$, but  $T_{BDD}$ stays higher.
The fact that Facets-To-Faces  is implemented in high-level Mathematica code, whereas {\tt BooleanConvert} is {\it hardwired}, is only part of the explanation. Fortunately according to 4.3.1 Facets-To-Faces can be parallelized. Thus, simply put, Facets-To-Faces can be accelerated by any factor $t$, provided one is fit to 'control' a network of $t$ colleagues who lend their PC's.

 {\bf 5.4} Not only is $R_{BC}$ always larger than $R$, the Mathematica command {\tt MemoryInUse} (whatever its units)  shows that   {\tt BooleanConvert} is also more memory-intensive than the Facets-To-Faces algorithm. For example, in a random instance of type $(50,200,20)$ the before/after measurements were ${\tt MemoryInUse}={\bf 307}'572'224$ and ${\tt MemoryInUse}={\bf 928}'179'088$ for Facets-To-Faces, but  ${\tt MemoryInUse}={\bf 307}'339'408$ and ${\tt MemoryInUse}={\bf 3'434}'044'152$ for {\tt BooleanConvert}.
 As to {\tt SatisfiabilityCount}, in the (2000,70,192)-example it also started with a modest ${\tt MemoryInUse}={\bf 62}'485'184$ but ended with a hefty ${\tt MemoryInUse}={\bf 5'698}'713'009$.
 This may be related to why ${\tt Timing[BooleanConvert]}$ and {\tt Timing[SatisfiabilityCount]} were not reliable: In the $(50,240,20)$ instance, say, the claim ${\tt Timing[BooleanConvert]=47sec}$ was contradicted by a hand-stopped time of 410 seconds. In the $(2000,70,192)$ instance the claim $\hfill$  {\tt Timing[SatisfiabilityCount]=157} was contradicted by a hand-stopped time of 785 seconds. In contrast, for the Facets-To-Faces algorithm  ${\tt Timing[...]}$ always matched the hand-stopped time. 
 
{\bf 5.5} From the fancy ESOP calculated by  Facets-To-Faces  (say in time $T$) one can compute all face-numbers in a fraction of $T$. This method may even beat e+rp+sub. Which method excels depends on the number and structure of facets and needs further investigation. For instance, e+rp+sub  was slightly faster on the (50,1000,10)-example (901 seconds) but much slower on the (2000,70,192)-example which was stopped after an hour.

\begin{tabular}{l|c|c|c|c||c|c|c||c|c|c|} 
& w & h & fs & $|{\cal S}{\cal C}|$ & $R$ & $R_{BC}$ & $R_{BDD}$& $T$ & $T_{BC}$ & $T_{BDD}$ \\ \hline
 &  &   &   &   &   &   & &  & \\ \hline
  $ $ & 50  & 10 &10 & $10^4$ & 39 & 200& 440    &0 & 0 & 0.3\\ \hline
   $ $ & 50  & 10 & 25 & $3\cdot 10^8$ & 285 & 5992& 62503    &0.03 & 0.06 & 15\\ \hline
   
  $ $ & 50  & 10 & 40 & $10^{13}$ & 429 & 220'432& 2'389'705    &0.1 & 2.4 & 139\\ \hline
  $ $ & 50  & 20 & 40 & $2\cdot 10^{13}$ & 46'473 & 4'988'955& 116'539'111    &5 & 63 & 8768\\ \hline
 $ $ & 50  & 30 & 10 & $3\cdot 10^{4}$ & 300 & 1007& 2662    &0.15 & 0.02 & 0.2\\ \hline
  $ $ & 50  & 30 & 25 & $10^{9}$ & 12'928 & 109'959& 1'263'264    &3 & 2 & 241\\ \hline
  $ $ & 50  & 30 & 40 & $3\cdot 10^{13}$ & 390'252 & 18'856'647&    &49 & 334 & \\ \hline
  $ $ & 50 & 40 & 30 & $4\cdot 10^{10}$ & 135'954 & 1'131'145 & 33'055'911  & 54 & 235& 4260\\ \hline
  $ $ & 50  & 60 & 25 & $2\cdot 10^{9}$ & 101'044 & 614'766& 6'236'949    &64 & 25 &995 \\ \hline
 $ $ & 50 & 60 & 30 & $6\cdot 10^{10}$ & 633'370 & 4'409'526 &   &225 &150 & \\ \hline
 $ $ & 50 & 80 & 25 & $3\cdot 10^{9}$ & 221'869 & 1'214'638 &    &194 &56 & \\ \hline
 $ $ & 50 & 100 & 10 & $9\cdot 10^{4}$ & 2249 & 5733 &13'783     &4 & 0.2& 8\\ \hline
 $ $ & 50 & 100 & 20 & $ 10^{8}$ & 66'606 & 247'749 &1'526'544    &50 & 6&384 \\ \hline
 $ $ & 50 & 240 & 20 & $2\cdot 10^{8}$ & 460'631 & 1'300'394 &    & 1420  & 410& \\ \hline 
 
$ $ & 50 & 300 & 10 & $3\cdot 10^{5}$ & 12'122 & 24'135 &    &59 & 0.6& \\ \hline
$ $ & 50 & 300 & 17 & $4\cdot 10^{7}$ & 204'177 & 515'639 &    &1228 & 17& \\ \hline
$ $ & 50 & 300 & 20 & $3\cdot 10^{8}$ & 722'394' & 1'938'347 &    &4362 & 121 & \\ \hline
$ $ & 50 & 300 & 26 & $3\cdot 10^{5}$ & 10'814'946 & 31'885'940 &   &45'998 & 1950& \\ \hline

$ $ & 50 & 1000 & 10 & $8\cdot 10^{5}$ & 61'982 & 101'269 &    &$ 1051 $ & 3& \\ \hline\hline

$ $ & 2000 & 70 & 50 & $9\cdot 10^{16}$ & 1523 & 44'786   &    & 4 & 65    & \\ \hline
$ $ & 2000 & 70 & 192 & $ 4\cdot 10^{59}$ & 70'551 &   &   & 99  &    & \\ \hline
$ $ & 2000 & 70 & 300 & $ 10^{92}$ & 707'518 &   &     & 1114  &   &    \\ \hline

\end{tabular}

{\sl Table 6: Facets-To-Faces versus {\tt BooleanConvert} and {\tt expr2bdd}}

\section{Two ways to enumerate ${\cal S}{\cal C}[k]$ from the facets of ${\cal S}{\cal C}$}

From among the four tasks listed in 1.1, only $(E_k)$ remains to be dealt with. We offer two methods.
Method 1  (in 6.1) is  faster,  but Method 2 (in 6.2) boasts a theoretic assessment.

 {\bf 6.1} In what follows any representation of ${\cal S}{\cal C}$ as disjoint union of 012e-rows  can be used as prerequisite for a compressed enumeration of
${\cal S}{\cal C}[k]$. For instance, the output of the Facets-To-Faces algorithm  in Table 5 would do, but for variation\footnote{This representation actually stems from a variant of Faces-To-Facets discussed in Section 8.} we chose to illustrate (see Table 7) our method on the representation ${\cal S}{\cal C}_1=\alpha_1\uplus \cdots\uplus \alpha_6$; this equality can be verified ad hoc\footnote{ Note that $|\alpha_1|+\cdots+|\alpha_6|=16+12+\cdots+2=52$ as it must. It thus suffices to show that each $\alpha_i$ is contained in $F_1\cup\cdots\cup F_5$, which is easy.}. 

Additionally to the e-wildcard we now need the {\it g-wildcard} $(g(t),...,g(t))$ which means 'exactly $t$ many 1's here'. Accordingly {\it 01g-rows}  are defined, e.g. $(1,0,g(2),g(2),g(2))$ is\\
 $\{(1,0,{\bf 1},{\bf 1},0),\ 
(1,0,{\bf 1},0,{\bf 1}),\ (1,0,0,{\bf 1},{\bf 1})\}$. Distinct g-wildcards within a 01g-row are distinguished by subscripts. Note that $t$ must be strictly smaller than the number of symbols $g(t)$ because $(g(3),g(3))$ is impossible, and instead of $(g(3),g(3),g(3))$ we stick to $(1,1,1)$.

{\bf 6.1.1} We now describe the {\it g-algorithm} which, given $\SC$ and $k\ge 0$, represents $\SC[k]$ as a disjoint union of 01g-rows. To fix ideas,
 let us target ${\cal SC}_1[3]$. The subset $\alpha_1[3]\s\alpha_1$ of all $3$-faces (=bitstrings of Hamming weight 3)  can  be written as $\gamma_1 = (g(3),g(3), g(3), 0, 0, 0, 0,0, g(3))$  (Table 7).
 Expressing $\alpha_2[3]$ similarly is a bit subtler. But writing $\alpha_2=(0,0,2,0,{\bf 1},0,{\bf 1},0,2)\uplus (0,0,2,0,{\bf g(1)},0,{\bf g(1)},0,2)$ one sees that the sets of 3-faces in the first and second part of $\alpha_2$ are $\gamma_{2,1}$ and $\gamma_{2,2}$ respectively.
 Likewise one verifies that $\alpha_3[3]=\gamma_3,\ \alpha_4[3]=\gamma_{4,1}\uplus\gamma_{4,2},\ \alpha_5[3]=\gamma_5,\
 \alpha_6[3]=\gamma_{6,1}\uplus\gamma_{6,2}$.

\begin{tabular}{l|c|c|c|c|c|c|c|c|c||c|} 
& 1 & 2 & 3 & 4 & 5 & 6 & 7 & 8 & 9 \\ \hline
 &  &   &   &   &   &   &  &   & \\ \hline
$ \alpha_1=$ & 2 & 2 & 2 & 0 & 0 & 0 & 0 & 0 & 2 & 16 \\ \hline
$\alpha_2=$ & 0 & 0 & 2 & 0 & $e$ & 0 & $e$ & 0 & 2 & 12\\ \hline

$\alpha_3=$ & 0 & 1 & 0 & 0 & 0 & 0 & 1 & 0 & 2 & 2 \\ \hline
$\alpha_4=$ & 0 & 0 & 2 & 0 & 0 & $e$ & 0 & $e$ & 2 & 12 \\ \hline

$\alpha_5=$ & 0 & 2 & 0 & 1 & 0 & 0 & 0 & 0 & 2 & 4\\ \hline
$\alpha_6=$ & 0 & e & 0 & e & 0 & 0 & 0 & 1 & 2 & 6 \\ \hline 
  &  &   &   &   &   &   &  &   & \\ \hline
 $\gamma_1= $ & g(3)  & g(3) & g(3) & 0 & 0 & 0 & 0 & 0 & g(3) & 4\\ \hline
 $\gamma_{2,1}= $ & 0 & 0 & g(1) & 0 & {\bf 1} & 0 & {\bf 1} & 0 & g(1) & 2\\ \hline
 $\gamma_{2,2}= $ & 0 & 0 & 1 & 0 & {\bf g(1)} & 0 & {\bf g(1)} & 0 & 1 & 2\\ \hline 
 $\gamma_3= $ & 0 & 1 & 0 & 0 & 0 & 0 & 1 & 0 & 1 & 1\\ \hline
 $\gamma_{4,1}= $ & 0 & 0 & g(1) & 0 & 0 & {\bf 1} & 0 & {\bf 1} & g(1) & 2\\ \hline
 $\gamma_{4,2}= $ & 0 & 0 & 1 & 0 & 0 & {\bf g(1)} & 0 & {\bf g(1)}& 1 & 2\\ \hline
 $\gamma_5= $ & 0 & 1 & 0 & 1 & 0 & 0 & 0 & 0 & 1 & 1\\ \hline
$\gamma_{6,1}= $ & 0 & {\bf 1} & 0 & {\bf 1} & 0 & 0 & 0 & 1 & 0 & 1\\ \hline
$\gamma_{6,2}= $ & 0 & {\bf g(1)} & 0 & {\bf g(1)} & 0 & 0 & 0 & 1 & 1 & 2\\ \hline
\end{tabular}

{\sl Table 7: Compressing ${\cal SC}_1[3]$ with the $g$-algorithm}

  What happens if, other than in Table 7, the $012e$-rows $\alpha$ that constitute ${\cal S}{\cal C}$ feature {\it several} $e$-wildcards per row? For instance if

(16) \quad $r = (e_1, e_1, e_1, e_1, e_1, \ e_2, e_2, e_2, \ e_3, e_3, e_3, \ e_4, e_4, \ e_5, e_5, 
\ e_6, e_6, \ 2, 1, 1, 0),$

 how can one represent $r[15]$ as disjoint union of preferably few $01g$-rows? Although even one-by-one enumeration of $r[15]$ is non-trivial (this type of task is solvable in output-linear time [W2, sec. 3.2]), proceeding differently one can actually get a compressed enumeration. This is carried out (on a dual example)  in a previous version of our article [arXiv:1812.02570v2, sec. 3.3.2], and it is fairly clear that matters generalize.
 
{\bf 6.1.2} More important than giving a formal proof of 'fairly clear'  is another issue. Suppose our target had been $\SC_1[4]$, not $\SC_1[3]$. Then $\alpha_5[4]=\emptyset$. Generally in the worst case a representation of $\SC$ by disjoint 012e-rows $\alpha_i$ might be such that $99\%$ of all rows $\alpha_i$ have $\alpha_i[k]=\emptyset$. Although this {\it empty-row-issue} prevents a (neat) theoretic assessment of the $g$-algorithm, this does not preclude a good performance in practise.

{\bf 6.2} Here we fine-tune the Second Naive Algorithm of 4.1.2 from outputting all faces to outputting all $k$-faces.

{\bf Theorem 2:} {\it Suppose the $h$ facets of the simplicial complex ${\cal S}{\cal C} \subseteq {\cal P}[w]$ are given. Then for any fixed $k \in [w]$ the $R$ many $k$-faces can be enumerated in time $O(Rhw^2)$.}

{\it Proof.} Starting with $(2,2, \cdots, 2) = {\cal P}[w]$  one maintains an oscillating (LIFO) stack of {\it $k$-feasible} $012$-rows $r$ (i.e. $r \cap {\cal S}{\cal C}[k] \neq \emptyset$) until the stack is emptied. The topmost row $r$ of the stack is always processed as follows. Let $r_0$ and $r_1$ be the rows obtained from $r$ by turning its first 2 to 0 and 1 respectively. (Here 'first' refers to some previosly fixed linear ordering of the indices $1,2,...,w$.) Row $r_0$ is $k$-feasible iff for at least one facet $F_i$ one has 

$(17)\quad ones(r_0) \subseteq F_i\ \hbox{\rm  and}\ |\mbox{\it ones}(r_0)| \leq k \leq |F_i\setminus zeros(r_0)|.$ 

 Likewise for $r_1$. At least one of $r_0$ and $r_1$ is $k$-feasible because $r$ is $k$-feasible and $r = r_0 \uplus r_1$. The feasible row(s) is (are) put back on the stack. That is unless (say) $r_0$ is a bitstring, i.e. twos$(r_0) = \emptyset$. In this case we found a $k$-face $r_0$, which is output.

As to the cost, creating $r_0,\ r_1$ from $r$ and recycling at least one of them to  the stack, costs $O(wh)$. Each output $k$-face has at most $w$ recycled ancestors. It follows that the overall cost is $O(R\cdot w\cdot wh)$. \quad $\square$

{\bf 6.3} Notwithstanding Theorem 2, in practise Method 2 (which like SNA in 4.1.3 suffers mediocre compression)
is often inferior to Method 1 (which laughs away its empty-row-issue). At this point the author may be forgiven for reflecting more broadly about one-by-one, compression, and optimization. As mentioned in 4.1.1  enumeration of a powerset (one-by-one) is non-trivial. Even more so enumerating all $k$-sets of a set [K, sec. 7.1.1.3]. Apart from the fun of it, arguably the only purpose of enumerating all objects of a given type, is to  find the best object (e.g. one or all $f$-minimal object(s) with respect to a target function $f$). Compression with multi-valued rows (be it 012e, 01g, or other kinds) serves that purpose better than one-by-one enumeration. If say $f(\{a_1,...,a_t\}):=a_1+\cdots+a_t$, then the $f$-minimal set $A$ within the 01g-row below is quickly found to be 
$A=\{2,4,7,9,10\}$ (for brevity $g_i:=g_i(1)$):

\begin{tabular}{c|c|c|c|c|c|c|c|c|c|c|c|c|} 
	 1 & 2 & 3 & 4 & 5 & 6 & 7 & 8 & 9 &10 &11 &12 &13 \\ \hline

	  0  & $g_1$ & 0 & 1 & 0 & $g_1$ & $g_2$ & $g_2$ & 1 &$g_3$ &$g_2$ &$g_3$ &$g_3$\\ \hline
	 	
\end{tabular}

More about the interplay of compression and optimization can be found in [W4].

\section{Assessing ${\cal S}{\cal C}$ from its minimal non-faces}

Our results on $(E),(E_k),(C),(C_{\forall k})$ will be adapted (in that order)  to the  situation where not the faces but  the minimal non-faces of a simplicial complex $\SC$ are given. For instance let $\SC$ be the family of all independent sets of a matroid. Then the facets are the bases and the minimal non-faces are the circuits of the matroid. 
  Dual to the $e$-widcard  the $n$-wildcard $(n,n,...,n)$ means 'at least one 0 here',  and {\it 012n-rows} are defined dually to 012e-rows.
Generally if $r$ is a $012n$-row with $\gamma: = |\mbox{twos}(r)|$ and with $s$ many $n$-wildcards of length $\delta_1, \cdots, \delta_s$ respectively, then 

(18) \quad $|r| = 2^\gamma \cdot (2^{\delta_1} -1) \cdot (2^{\delta_2} -1) \cdots (2^{\delta_s} -1)$.

 If ${\cal G}\s{\cal P}(W)$ is any hypergraph, then $X\s W$ is a  ${\cal G}$-{\it noncover} if $X\not \supseteq G$ for all $G\in {\cal G}$.
The {\it (noncover) n-algorithm} of [W1] displays the set ${\cal N}({\cal G})$ of all ${\cal G}$-noncovers
as a disjoint union of 012n-rows. (More  details about the $n$-algorithm follow in the proof of Theorem 3.)

{\bf 7.1 } Here we settle (E). Suppose  that ${\cal SC}_1$ was given not by its facets listed in (1), but by its minimal nonfaces, which are these:

(19) \quad $G_1 = \{1, 5\}, \ G_2 = \{2, 5\}, \ G_3 = \{1, 7\}, \ G_4 = \{2, 3, 7\}, \ G_5  = \{1, 6\}, \ G_6= \{2, 6\}$,

\hspace*{.9cm} $G_7 = \{ 5, 6\}, \ G_8 = \{6,7\}, \ G_9 = \{1, 8\}, \ G_{10} = \{5, 8\}, \ G_{11} =  \{7, 8\}, \ G_{12}=\{1, 4\}$,

\hspace*{.9cm} $G_{13} = \{3, 4\}, \ G_{14} = \{4, 5\}, \ G_{15} = \{4, 6\}, \ G_{16} = \{4, 7\}, \ G_{17} = \{2, 3, 8\}$.

For instance, $G_{17}$ is not a subset of any $F_i$ in (1), but each 2-element subset of $G_{17}$ is contained in some $F_i$. Hence $G_{17}$ is a minimal non-face. Generally, let ${\cal S}{\cal C}$ be given by its minimal nonfaces $G_1, G_2,..$, and so forth. It then holds that

(20) \quad $X \in {\cal S}{\cal C} \ \Leftrightarrow \ (\forall i) X \not\supseteq G_i \ \Leftrightarrow \ (\forall i)(X^c \cap G_i \neq \emptyset)$.

 From the first equivalence it follows that ${\cal SC}_1={\cal N}({\cal G}_1)$ where ${\cal G}_1=\{G_1, \cdots, G_{17}\}$. Applying the  $n$-algorithm  to ${\cal G}_1$ delivers ${\cal SC}_1$ as a disjoint union of the $012n$-rows  $r_1,\cdots, r_7$ in Table 8. (Incidently only $r_2$ is a proper 012n-row.)

\begin{tabular}{l|c|c|c|c|c|c|c|c|c|} 
	& 1 & 2 & 3 & 4 & 5 & 6 & 7 & 8 & 9 \\ \hline
	&  &   &   &   &   &   &  &   & \\ \hline
	$r_1 =$ & 0  & 2 & 0 & 2 & 0 & 0 & 0 & 2 & 2 \\ \hline
	$r_2=$ & 0 & $n$ & 1 & 0 & 0 & 0 & 0 & $n$ & 2\\ \hline
	$r_3=$ & 1 & 2 & 2 & 0 & 0 & 0 & 0 & 0 & 2 \\ \hline 
	$r_4=$ & 0 & 0 & 2 & 0 & 0 & 1 & 0 & 2 & 2 \\ \hline
	$r_5=$ & 0 & 0 & 2 & 0 & 0 & 0 & 1 & 0 & 2 \\ \hline
	$r_6=$ & 0 & 1 & 0 & 0 & 0 & 0 & 1 & 0& 2\\ \hline
	$r_7=$ & 0 & 0 & 2 & 0 & 1 & 0 & 2 & 0 & 2\\ \hline
\end{tabular}

{\it Table 8: Compressing ${\cal SC}_1$ with the noncover $n$-algorithm}

{\bf Theorem 3:} {\it Assume the $h$ minimal non-faces of the simplicial complex ${\cal S}{\cal C} \subseteq {\cal P}[w]$ are known. Then ${\cal S}{\cal C}$ can be represented as a disjoint union of $R$ many $012n$-rows in polynomial total time $O(Rh^2w^2)$.}

{\it Proof.} The minimal non-faces $G_i$ in (17) suggest to view ${\cal SC}_1$  (or any ${\cal S}{\cal C})$ as the model set $\mbox{Mod}(\varphi_1) : = \{u \in \{0,1\}^9 : \varphi_1(u) = 1\}$ of the Boolean function\footnote{Because of $Mod(\varphi_1)=Mod(\psi_1)=\SC_1$ we have $\varphi_1=\psi_1$, despite appearances.}

(21) \quad $\varphi_1 (x_1, \cdots, x_9) : = (\ol{x}_1 \vee \ol{x}_5) \wedge (\ol{x}_2 \vee \ol{x}_5) \wedge \cdots \wedge (\ol{x}_2 \vee \ol{x}_3 \vee \ol{x}_8)$

This is a {\it Horn-CNF} since each clause has at most one positive literal (in fact none). Generally, if $\varphi : \{0,1\}^w \ra \{0,1\}$ is a Horn-CNF with $h$ clauses then the Horn-$n$-algorithm of [W1, Cor.6] enumerates $\mbox{Mod}(\varphi)$ as a union of $R$ many disjoint $012n$-rows in total polynomial time $O(Rh^2w^2)$. \quad $\square$

When the Horn-CNF has only negative clauses, the Horn $n$-algorithm simplifies and was called 'noncover $n$-algorithm' in [W1].  The impression from (20) that  the noncover n-algorithm is related to the transversal e-algorithm is justified; in fact a moment's thought reveals that upon switching the roles of 0 and 1 the n-algorithm {\it becomes} the e-algorithm, and vice versa.

One application of Theorem 3 was alluded to in Section 1.3: From a knowledge of all minimal infrequent sets, one can compress the simplicial complex of all frequent sets. 

{\bf 7.1.2} As to problem $(E_k)$, i.e. the enumeration of all $k$-faces from the minimal non-faces, this can be handled by applying the (dual) $g$-algorithm to the individual 012n-rows in Table 8.  Trouble is, as in 6.1 this does not yield a polynomial total time algorithm because of the empty-row-issue. It remains an open question whether the analogon of Theorem 2 holds. More precisely by 'analogon' we mean the statement that ensues from Theorem 2 when the part 'the $h$ facets' is replaced by 'the $h$ minimal non-faces'. The problem is that (17) does not translate smoothly from facets $F_i$ to minimal non-faces $G_i$.

{\bf 7.2} As to the counting problem (C), the cardinality of ${\cal SC}_1$ is readily obtained from Table 8:

(22)\qquad  $|{\cal SC}_1|= |r_1| + \cdots + |r_7|=16+6+8+8+4+2+8=52.$

As to problem $(C_{\forall k})$, each face-number $N_k$ of ${\cal SC}_1$ can  be calculated  from Table 8 by matching each 012n-row with some auxiliary polynomial, akin to 3.3. We hence call this method {\it n+rp+sub}.

{\bf 7.3 }  {\it Hypergraph Dualization (HD)} is the task to calculate the set ${\cal M}{\cal T}({\cal H})$ of all minimal transversals of a hypergraph  ${\cal H} \subseteq  {\cal P}[w]$. This has plenty applications. As to HD  in the present situation, let $\SC$ be a simplicial complex. Then by (20), the  complements of its facets $F_i$'s are exactly the minimal transversals of its minimal non-faces $G_i$'s, and vice versa. Thus if HD was easy, one could  switch back and forth between the $F_i$'s and $G_i$'s at one's convenience; for instance discarding the seventeen $G_i$'s in (19) in favor of the five $F_i$'s in (1).

Unfortunately HD is hard. Despite partial successes it remains an open
 problem whether HD can be solved in polynomial total time. We stress that the e-algorithm computes the set ${\cal T}({\cal H})$ of {\it all} transversals. Extra work\footnote{This can actually be done, not in total polynomial time, but whilst maintaining compression to some degree; this is work in progress, arXiv:2008.08996. In one special case this worked particularly well: If ${\cal H}$ is the set of all minimal cutsets of a graph $G$, then ${\cal T}({\cal H})$ is the set of all connected edge-sets. In arXiv:2002.09707 it is shown how the
family  ${\cal M}{\cal T}({\cal H})$ of all trees can be compressed.    } is required to 'sieve' ${\cal M}{\cal T}({\cal H})$ from ${\cal T}({\cal H})$.

\section{Can one go from simplicial complexes to  general DNFs ?}

Suppose ${\cal S}{\cal C}$ has facets $F_1$ to $F_h$, and by induction we have obtained for some $t \in [h-1]$ a type (18) representation.
In Section 8 we  handle the newcomer $012$-row $r:={\cal P}(F_{t+1})$ in dual fashion:

(23) \quad ${\cal P}(F_1) \cup \cdots \cup {\cal P}(F_{t+1}) = \rho_1 \uplus \cdots \uplus \rho_s\uplus (r\setminus (\rho_1 \uplus \cdots \uplus \rho_s)).$

We keep the notation $r_i=
{\cal P}(F_i)$ for $i\le 5$, and refer to Table 9 for the definition of $r'_i\ (i\ge 6)$. Furthermore, put say  $A\setminus B\setminus C\setminus D:=
((A\setminus B)\setminus C)\setminus D$. Based on (23) our {\it Tentative Facets-To-Faces} algorithm  proceeds as follows in our toy example ${\cal S}{\cal C}_1={\cal P}(F_1)\cup\cdots\uplus{\cal P}(F_5)$:

$\begin{array}{rll}
r_1 \cup r_2 = r_1\uplus (r_2 \setminus r_1)      & =: & r_1 \uplus r'_6 \\
\\
r_1 \uplus r'_6\uplus(r_3\setminus (r_1 \uplus r'_6)) = r_1 \uplus r'_6\uplus(r_3\setminus r_1 \setminus r'_6) &=:& r_1 \uplus r'_6\uplus r'_7  \\
\\
r_1 \uplus r'_6\uplus r'_7 \uplus (r_4\setminus r_1\setminus r'_6\setminus r'_7)         & =:& r_1 \uplus r'_6\uplus r'_7\uplus r'_8 \\
\\
r_1 \uplus r'_6\uplus r'_7\uplus r'_8\uplus (r_5\setminus r_1\setminus r'_6\setminus r'_7\setminus r'_8)  & =: & r_1 \uplus r'_6
\uplus r'_7\uplus r'_8\uplus \rho'_1\uplus \rho'_2\\

\end{array}$

Note that $r_4\setminus r_1$ is disjoint from $r'_6$ and $r'_7$, and hence $r_4\setminus r_1\setminus r'_6\setminus r'_7=r_4\setminus r_1=:r'_8$. Likewise
$r_5\setminus r_1$ being disjoint from $r'_6$ and $r'_7$ implies $r_5\setminus r_1\setminus r'_6\setminus r'_7=r_5\setminus r_1=:\rho'$. 
The detachment of $\rho_8'$ from $r_8'$ is of type $012e\setminus 012e$ as opposed to $012e\setminus 02$ in Section 4. Before we look at    
type $012e\setminus 012e$ detachments more systematically
we argue ad hoc as follows. Since $\rho'\cap r'_8,\ \rho'_1,\ \rho'_2$ are contained in $\rho'$, and are mutually disjoint, and their cardinalities sum up to $2+4+6=|\rho'|$, it follows that $\rho'\setminus r'_8=\rho'_1\uplus \rho'_2$.  One checks that

(24)\quad $|r_1|+|r'_6|+|r'_7|+|r'_8|+|\rho'_1|+|\rho'_2|=16+12+2+12+4+6=52,$

which matches the cardinality $|{\cal S}{\cal C}_1|$ (which we previously derived in various ways).

\begin{tabular}{r|c|c|c|c|c|c|c|c|c|l} 
& 1 & 2 & 3 & 4 & 5 & 6 & 7 & 8 & 9 & \\ \hline
&   &   &   &   &   &   &   &   &   & \\ \hline
$r_1=$ & 2 & 2 & 2 & 0 & 0 & 0 & 0 & 0 & 2 & 16 \\ \hline
$r_2=$ & 0 & 0 & 2 & 0 & 2 & 0 & 2 & 0 & 2 \\ \hline
$r_3=$ & 0 & 2 & 0 & 0 & 0 & 0 & 2 & 0 & 2 \\ \hline
$r_4=$ & 0 & 0 & 2 & 0 & 0 & 2 & 0 & 2 & 2 \\ \hline
$r_5=$ & 0 & 2 & 0 & 2 & 0 & 0 & 0 & 2 & 2 \\ \hline
$r'_6=$ & 0 & 0 & 2 & 0 & $e$ & 0 & $e$ & 0 & 2 & 12\\ \hline
$r_3\setminus r_1=$ & 0 & 2 & 0 & 0 & 0 & 0 & 1 & 0 & 2  \\ \hline
$r'_7=$ & 0 & 1 & 0 & 0 & 0 & 0 & 1 & 0 & 2 & 2 \\ \hline
$r'_8=r_4\setminus r_1=$ & 0 & 0 & 2 & 0 & 0 & $e$ & 0 & $e$ & 2 & 12 \\ \hline
$\rho'=r_5\setminus r_1=$ & 0 & 2 & 0 & $e$ & 0 & 0 & 0 & $e$ & 2 \\ \hline
$\rho'\cap r'_8=$ & 0 & 0 & 0 & 0 & 0 & 0 & 0 & 1 & 2 \\ \hline
$\rho'_1=$ & 0 & 2 & 0 & 1 & 0 & 0 & 0 & 0 & 2 & 4\\ \hline
$\rho'_2=$ & 0 & e & 0 & e & 0 & 0 & 0 & 1 & 2 & 6 \\ \hline

\end{tabular}

{\sl Table 9: Compressing ${\cal S}{\cal C}_1$ with a Tentative Facets-To-Faces algorithm }

{\bf 8.1} We saw that initial $02\setminus 02$ detachments can quickly 'deteriorate' to $012e\setminus 012e$ detachments such as $\rho'\setminus r_8'$. While $\rho'\setminus r_8'$ was handled ad hoc, let us now dig  deeper.
Namely, by definition $mm..m$ means 'at least one 1 and at least one 0 here'. Let $\rho$ and $r$ be as in Table 10. With our new wildcard the
row difference $\rho\setminus r$ can be neatly expressed as $\rho_1\uplus\rho_2$. Indeed, clearly $\rho_1\uplus\rho_2\subseteq \rho\setminus r$. If there was $x\not\in \rho\setminus r$ with $x\not\in \rho_1\uplus\rho_2$ then   $x_4=x_5=x_6=0$ leads to the contradiction ${\bf x}\in(2,2,1,0,0,0)\subseteq r$.

\includegraphics[scale=1.1]{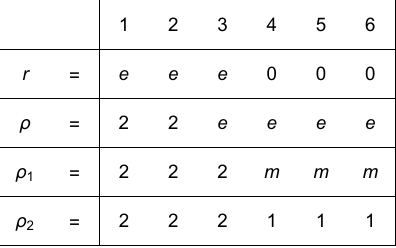}

{\sl Table 10: Using the $mm...m$ wildcard to recompress $\rho\setminus r$}

As appealing as this may look, the downside is that embracing $012men$-rows forces us to cope with detachments of type $012men\setminus 012men$.  Table 11 must suffice as indication that things do not get out of hand. The verification that indeed $\rho\setminus r=\rho_1\uplus\cdots\uplus\rho_{13}$ is left to the dedicated reader.

\begin{tabular}{l|c|c|c|c|c|c|c|c|c|c|} \hline
$\rho=$ & $e$ & $n$ & $m$ & 2 & $n$ & $e$ & $m$ & $m$ & $e$ & $n$\\ \hline 
$r=$ & 1 & 1 & $e_1$ & $e_1$ & $e_2$ & $e_2$ & 0 & $n$ & $n$ & 2\\ \hline
&     &    &     &         &      &        &    &    &    & \\ \hline
$\rho_1 =$ & ${\bf 0}$ & $n$ & $m$ & 2 & $n$ & $e$ & $m$ & $m$ & $e$ & $n$\\ \hline
$\rho_2 =$ & ${\bf 1}$ & ${\bf 0}$ & $m$ & 2 & 2 & 2 & $m$ & $m$ & $2$ & $2$\\ \hline
$\rho_3 =$ & ${\bf 1}$ & ${\bf 1}$ & ${\bf 0}$ & ${\bf 0}$ & $n$ & $2$ & $e'$ & $e'$ & $2$ & $n$\\ \hline
$\rho_4 =$ & ${\bf 1}$ & ${\bf 1}$ & ${\bf 1}$ & ${\bf 2}$ & ${\bf 0}$ & ${\bf 0}$ & $n'$ & $n'$ & $2$ & $2$\\ \hline
$\rho_5 =$ & ${\bf 1}$ & ${\bf 1}$ & ${\bf 0}$ & ${\bf 1}$ & ${\bf 0}$ & ${\bf 0}$ & $e'$ & $e'$ & $2$ & $2$\\ \hline
&     &    &     &         &      &        &    &    &    & \\ \hline
$\rho_6=$ & 1 & 1 & 1 & 2 & 1& 2 & ${\bf 1}$ & 0 & 2 & 0\\ \hline
$\rho_7=$ & 1 & 1 & 1 & 2 & 1 & 2 & ${\bf 0}$ & ${\bf 1}$ & ${\bf 1}$ & 0\\ \hline
$\rho_8=$ & 1 & 1 & 1 & 2 & 0 & 1 & ${\bf 1}$ & 0 & 2 & 2\\ \hline
$\rho_9=$ & 1 & 1 & 1 & 2 & 0 & 1 & ${\bf 0}$ & ${\bf 1}$ & ${\bf 1}$ & 2\\ \hline
$\rho_{10}$ & 1 & 1 & 0 & 1 & 1 & 2 & ${\bf 1}$ & 2 & 2 & 0\\ \hline
$\rho_{11}=$ & 1 & 1 & 0 & 1 & 1 & 2 & ${\bf 0}$ & ${\bf 1}$ & ${\bf 1}$ & 0\\ \hline
$\rho_{12}=$ & 1 & 1 & 0 & 1 & 0 & 1 & ${\bf 1}$ & 2 & 2 & 2\\ \hline
$\rho_{13}=$ & 1 & 1 & 0 & 1 & 0 & 1 & ${\bf 0}$ & ${\bf 1}$ & ${\bf 1}$ & 2\\ \hline
\end{tabular}

{\sl Table 11 : Recompression of a set difference $\rho \setminus r$ of type $(012men) \setminus (012en)$ }

Once $012men\setminus 012men$ detachments are mastered, {\it any} DNF can  be 
transformed to a fancy
 ESOP that uses $012men$-rows. Given a CNF instead of a DNF, a wholly different method  to transform the CNF to a fancy ESOP (using 012n-rows) is presented in [W4].

\section*{References}

\begin{enumerate}

\item[{[BM]}] J.D. Boissonnat, C. Maria, The simplex-tree: An efficient data structure for general simplicial complexes, Algorithmica 70 (2014) 406-427.
\item[{[BN]}] M.O. Ball, G.L. Nemhauser, Matroids and reliability analysis problem, Math. of Oper. Res. 4 (1979) 132-143.
\item[{[DKM]}] A.M. Duval, J. Klivans, J.L. Martin, The partitionability conjecture, Notices of the AMS 64 (2017) 117-122.
\item[{[FR]}] K. Fukuda, V. Rosta, Combinatorial face enumeration in convex polytopes, Computational Geometry 4 (1994) 191-198.
\item[{[GO]}] B. Ganter, S. Obiedkov, Conceptual Exploration, Springer 2016.
\item[{[KP]}] V. Kaibel, M.E. Pfetsch, Computing the face lattice of a polytype from its vertex-facet incidences, Computational Geometry 23 (2002) 281-290.
\item[{[K]}] D. Knuth, The Art of Computer Programming, Volume 4A, Addison-Wesley 2011.
\item[{[V]}] L.G. Valiant, The complexity of enumeration and reliability problems, SIAM J. Comput. 8 (1979) 410-421.
\item[{[W1]}] M. Wild, Compactly generating all satisfying truth assignments of a Horn formula, J. Satisf. Boolean Model. Comput. 8 (2012) 63-82.

\item[{[W2]}] M. Wild, Counting or producing all fixed cardinality transversals, Algorithmica 69 (2014) 117-129.
\item[{[W3]}] M. Wild, The joy of implications, aka pure Horn formulas: Mainly a survey, Theoretical Computer Science 658 (2017) 264-292.
\item[{[W4]}] M. Wild, Compression with wildcards: From CNFs to orthogonal DNFs by imposing the clauses one-by-one, to appear in The Computer Journal.
\item[{[WK]}] X. Wu, V. Kumar, The top ten algorithms in data mining, Chapman and Hall 2009.
.

\end{enumerate}

\end{document}